\documentclass[aps,prl,showpacs,notitlepage,twocolumn,superscriptaddress,nofootinbib,preprintnumbers]{revtex4-2}
\usepackage{bbm}
\usepackage{mathrsfs}
\usepackage{epsfig}
\usepackage{soul,xcolor}
\usepackage{graphicx}
\usepackage{amsfonts}
\usepackage{amsthm}
\usepackage[figuresright]{rotating}
\usepackage{amssymb}
\usepackage{amsmath}
\usepackage{dcolumn}
\usepackage{physics}
\usepackage{float}
\usepackage{bm}
\usepackage{physics}
\usepackage{verbatim}
\usepackage{braket}
\usepackage[normalem]{ulem}
\usepackage[ruled,vlined,linesnumbered]{algorithm2e}
\usepackage{setspace}
\usepackage{lipsum}
\usepackage[colorlinks,linkcolor=blue,anchorcolor=blue,citecolor=blue,urlcolor=blue]{hyperref}

\begin{document}

\title{Variational Neural-Network Ansatz for Continuum Quantum Field Theory}
\author{John M. Martyn}
\affiliation{Center for Theoretical Physics, Massachusetts Institute of Technology, Cambridge, MA 02139, USA}
\affiliation{The NSF AI Institute for Artificial Intelligence and Fundamental Interactions}
\author{Khadijeh Najafi}
\affiliation{IBM Quantum, IBM T.J. Watson Research Center, Yorktown Heights, NY 10598 USA}
\affiliation{MIT-IBM Watson AI Lab,  Cambridge, MA 02142, USA}
\author{Di Luo}
\affiliation{Center for Theoretical Physics, Massachusetts Institute of Technology, Cambridge, MA 02139, USA}
\affiliation{The NSF AI Institute for Artificial Intelligence and Fundamental Interactions}
\affiliation{Department of Physics, Harvard University, Cambridge, MA 02138, USA}

\begin{abstract}
Physicists dating back to Feynman have lamented the difficulties of applying the variational principle to quantum field theories. In non-relativistic quantum field theories, the challenge is to parameterize and optimize over the infinitely many $n$-particle wave functions comprising the state's Fock space representation. Here we approach this problem by introducing \emph{neural-network quantum field states}, a deep learning ansatz that enables application of the variational principle to non-relativistic quantum field theories in the continuum. Our ansatz uses the Deep Sets neural network architecture to simultaneously parameterize all of the $n$-particle wave functions comprising a quantum field state. We employ our ansatz to approximate ground states of various field theories, including an inhomogeneous system and a system with long-range interactions, thus demonstrating a powerful new tool for probing quantum field theories. 
\end{abstract}

\preprint{MIT-CTP/5491}
\maketitle

\emph{Introduction.}---
It is notoriously challenging to solve an interacting quantum field theory (QFT), with analytical solutions limited to mean field systems and few exactly solvable models. While perturbation theory is useful for weakly-interacting field theories, much important physics lies in the non-perturbative regime, such as in quantum chromodynamics. Although lattice field theory is a well-established method for studying QFT, it can be computationally expensive to reach the continuum limit~\cite{rothe2012lattice}. It is therefore an ongoing quest to develop new methods for solving and simulating QFTs in the continuum.

An alternative approach to probing QFT is through applying the variational principle to field theory. However, as pointed out by Feynman~\cite{feynman1988difficulties}, a significant barrier to this method is constructing a variational ansatz for a quantum field state whose expectation values can be efficiently computed and optimized. Ref.~\cite{Verstraete2010} took a key step in this direction by extending the matrix product state (MPS) from lattice models to field theories through the development of the continuous matrix product state (cMPS), which can describe low energy states of 1D non-relativistic QFTs~\cite{Haegeman_2013_calculus}. However, unlike MPS on the lattice, cMPSs suffer substantial drawbacks in their optimization and applicability. While cMPS can be efficiently optimized in translation-invariant systems~\cite{Ganahl_2017}, in inhomogenous settings the direct optimization of cMPS is prone to numerical instabilities~\cite{Haegeman_2017, PhysRevLett_Tuybens}, and algorithms to better adapt cMPS to these settings ultimately rely on discretizations and/or interpolations~\cite{Ganahl_2017_Continuous, Ganahl2018PhRvB, PhysRevLett_Tuybens}. Moreover, approximation techniques are needed to apply cMPS to systems with generic long-range interactions~\cite{Lukin_2022, Rincon_2015_Lieb}, leaving cMPS often restricted to contact interactions. Furthermore, extensions of cMPS to higher spatial dimensions are susceptible to UV divergences and thus limited in application~\cite{Tilloy2019, Jennings_2015, Karanikolaou_2021, Shachar_2022}.

Meanwhile, the striking success of machine learning in physically-motivated contexts~\cite{Lin_2017} has recently inspired the application of neural networks to problems in quantum physics, leading to a new variational ansatz known as a \emph{neural-network quantum state} (NQS)~\cite{Carleo2017}. This ansatz parameterizes a wave function by neural networks, which are optimized over to approximate a state of interest~\cite{melko2019restricted, Choo_2018, torlai2018neural} or even simulate real-time dynamics~\cite{gutierrez2020real, Schmitt_2020, Vicentini_2019, Yoshioka_2019, quantum_circuit, Hartmann_2019, Nagy_2019, luo_gauge_inv, luo_povm, Reh_2021}. These methods are justified by recent proofs that NQS can efficiently model highly entangled states and subsume tensor networks in their expressivity~\cite{Deng_2017, Chen_2018, gao2017efficient, Glasser_2018, Levine_2019, sharir2021neural}.

However, NQS have not been generalized to QFT in the continuum, which would necessitate an efficient parameterization of the infinitely many $n$-particle wave functions comprising a quantum field state's Fock space representation. Prominent instances of NQS in continuous space have only studied systems with a fixed number of particles~\cite{Pescia_2022, wilson2022wave, Lovato_2022, Gnech_2021, Adams_2021, Pfau_2020, Hermann_2020}, and have not yet made the connection to QFT with variable particle number. Applications of machine learning to QFT have only studied lattice field theory, most notably Monte Carlo sampling in the Lagrangian formulation~\cite{PhysRevD.100.034515, PhysRevLett.126.032001, Albergo_2021, Kanwar_2020} and simulations in the Hamiltonian formulation~\cite{luo_gauge_inv,luo_u1,luo2021gauge}, rather than applying the variational principle directly to continuum QFT.

In this Letter, we fill this gap by developing \emph{neural-network quantum field states} (NQFS), extending the range of NQS to field theories. In this introductory work, we focus on non-relativistic QFT of bosons, while the approach we introduce opens up the opportunity to study generic field theories. Working directly in the continuum, we model a quantum field state by exploiting its Fock space representation as a superposition of $n$-particle wave functions. We adopt permutation invariant methods from deep learning -- specifically, the Deep Sets architecture~\cite{2017_Zaheer} -- to simultaneously parameterize the infinitely many $n$-particle wave functions by a finite number of neural networks. We further apply an algorithm for variational Monte Carlo (VMC) in Fock space, that enables the estimation and optimization of the energy of a quantum field state. The merit of NQFS lies in its applicability to various field theories, such as inhomogeneous systems and systems with long-range interactions, including both periodic and closed boundary conditions. We demonstrate these properties by employing NQFS to approximate the ground states of 1D QFTs, benchmarking on the Lieb-Liniger model, the Calogero-Sutherland model, and a regularized Klein-Gordon model.

\emph{Neural-Network Quantum Field States (NQFS).}---
We aim to approximate the ground state of a bosonic QFT, whereupon second quantization particles are created and annihilated by operators $\hat{\psi}^\dag(x)$ and $\hat{\psi}(x)$, respectively, which obey the commutation relation $[\hat{\psi}(x), \hat{\psi}^\dag(x')] = \delta(x-x')$. Focusing on non-relativistic systems in 1D, the corresponding quantum field states live in a Fock space that is a direct sum over all symmetrized $n$-particle Hilbert spaces, such that an arbitrary state may be expressed as a superposition of permutation invariant, unnormalized, $n$-particle wave functions $\varphi_n(\textbf{x}_n)$:
\begin{equation}
\begin{aligned}
    |\Psi\rangle = \bigoplus_{n=0}^\infty |\varphi_n\rangle = \bigoplus_{n=0}^\infty \int d^nx \ \varphi_n(\textbf{x}_n) |\textbf{x}_n\rangle.
\end{aligned}
\end{equation}
Here we use the shorthand $\textbf{x}_n = (x_1, x_2, \dots, x_n)$ to denote a vector containing the positions of $n$ particles, and employ the $n$-particle basis $|\textbf{x}_n \rangle :=  \frac{1}{\sqrt{n!}} \hat{\psi}(x_1)^\dag \hat{\psi}(x_2)^\dag \dots \hat{\psi}(x_n)^\dag |\Omega\rangle$, where $|\Omega \rangle$ is the vacuum.

We introduce NQFS as a variational ansatz to parameterize such a quantum field state with neural networks -- more specifically, to parameterize each of its $n$-particle wave functions with a common neural network architecture. This imposes the following constraints on the architecture. First, the architecture must be permutation invariant in order to model bosonic wave functions. And more stringently, to model wave functions of arbitrary particle number, the architecture must also be \emph{variadic}: able to accept an arbitrary number of arguments (i.e., particle positions $\textbf{x}_n = (x_1, x_2, \dots , x_n)$ for arbitrary $n$). 

Remarkably, we can realize both of these properties with the Deep Sets neural network architecture, which is both permutation invariant and variadic. Ref.~\cite{2017_Zaheer} introduces the Deep Sets architecture to model a permutation invariant function on a set $X$, and proves that any such function $f(X)$ can be decomposed as
\begin{equation}\label{eq:DeepSets_decomposition}
    f(X) = \rho\Big(\sum_{x \in X }\phi(x)\Big),
\end{equation}
for appropriately chosen constituent functions $\rho$ and $\phi$. The function $\phi$ maps the inputs into a feature space of possibly higher dimension, the sum aggregates the embedded inputs in a permutation invariant manner, and $\rho$ adds the correlations necessary to output the function of interest. This decomposition is clearly permutation invariant as each $\phi(x)$ is summed independently; it is also variadic because an arbitrary number of arguments may be included in the sum. 

\begin{figure}[tbp]
    \includegraphics[width=0.49\textwidth]{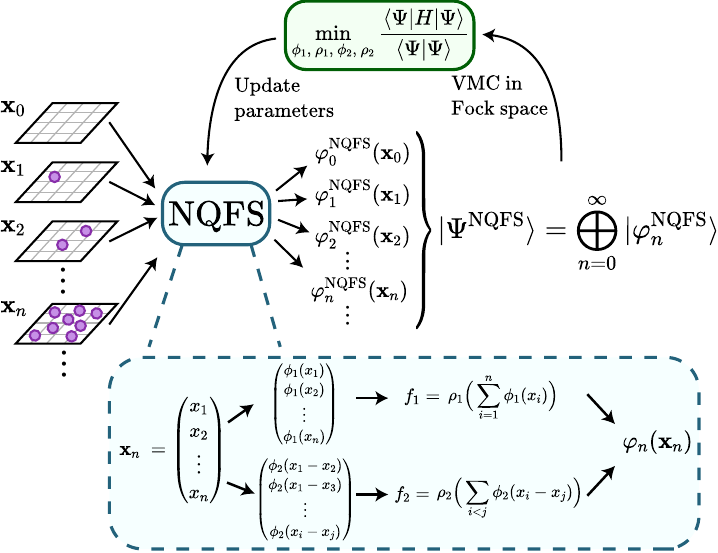}
    \caption{Diagram of a NQFS: particle positions $\textbf{x}_n = (x_1, x_2, \dots, x_n)$ are input to evaluate the corresponding $n$-particle wave functions $\varphi^{\text{NQFS}}_n(\textbf{x}_n)$, of which the NQFS $|\Psi^{\text{NQFS}}\rangle$ is modelled as a direct sum. Computation of $\varphi^{\text{NQFS}}_n(\textbf{x}_n)$ is shown in the inset as a product of two Deep Sets, $f_1$ and $f_2$. Optimization of the NQFS is illustrated in the feedback loop; the variational parameters are updated via VMC in Fock space to minimize the energy. \vspace{-14px}}
    \label{fig:NQFS_Diagram}
\end{figure}

In the setting of machine learning, this decomposition is used to learn permutation invariant functions by parameterizing $\rho$ and $\phi$ as deep neural networks~\cite{2017_Zaheer}. This approach has recently been used in NQS contexts to model ground states of atomic nuclei~\cite{Lovato_2022, Gnech_2021, Adams_2021} and bosons at fixed particle number~\cite{Pescia_2022}. Building on this prior work, here we are the first to leverage Deep Sets to develop a neural network ansatz for continuum QFT, with an arbitrary number of particles. 

For the construction of a NQFS, denoted $|\Psi^\text{NQFS} \rangle$, we parameterize each $n$-particle wave function as a product of two Deep Sets -- one for particle positions $\{x_i\}_{i=1}^n$, and the other for particle separations $\{x_i - x_j\}_{i<j}$:\footnote{While in principle, $f_1$ suffices to model an arbitrary state as per Ref.~\cite{2017_Zaheer}, we found it helpful to also include $f_2$ to better capture pairwise interactions.}
\begin{equation}\label{eq:NQFS_definition}
\begin{aligned}
    &|\Psi^{\text{NQFS}}\rangle = \bigoplus_{n=0}^\infty \int d^nx \ \varphi_n^\text{NQFS}(\textbf{x}_n) |\textbf{x}_n\rangle, \\
    &\varphi_n^\text{NQFS}(\textbf{x}_n) = \frac{1}{L^{n/2}} \cdot f_1\big( \{x_i\}_{i=1}^n \big) \cdot f_2\big( \{x_i-x_j\}_{i < j} \big),
\end{aligned}
\end{equation}
where the factor of $\tfrac{1}{L^{n/2}}$ is included for dimensional consistency. The constituent functions of the Deep Sets $f_1$ and $f_2$ ($\rho_1, \phi_1$ and $\rho_2, \phi_2$, respectively) are parameterized as feedforward neural networks, which capture the global properties and pairwise correlations of the state, respectively. Fundamentally, the variadic property of Deep Sets enables us to parameterize \emph{infinitely many} $n$-particle wave functions with a \emph{finite} number of neural networks, and thus represent an arbitrary state in Fock space. We provide a graphical illustration of a NQFS in Fig.~\ref{fig:NQFS_Diagram}. 

We employ feature embeddings on inputs to the neural networks. In a closed system with hard walls at $x=0$ and $x=L$, we use the embedding $x_i \mapsto (\frac{x_i}{L}, \ 1-\frac{x_i}{L})$ for $f_1$, and $(x_i-x_j) \mapsto \big(\frac{x_i-x_j}{L}\big)^2$ for $f_2$ (note that this embedding must be even in order to achieve permutation invariance). For a periodic system, we reflect this periodicity with the embedding $x_i \mapsto \big(\sin(\frac{2\pi }{L} x_i ) ,\ \cos(\frac{2\pi }{L} x_i)\big)$ for $f_1$, and $(x_i-x_j) \mapsto \cos(\frac{2\pi }{L}(x_i-x_j))$ for $f_2$.

\emph{VMC in Fock Space.}---
Our approach to ground state approximation employs the variational principle -- that the energy of an arbitrary quantum state is greater than or equal to the ground state energy. To estimate and optimize the energy of $|\Psi^{\text{NQFS}}\rangle$, we generalize traditional VMC to \emph{VMC in Fock space}. As we illustrate in the Supplementary Material, the energy of a quantum field state $|\Psi \rangle$ can be expressed as an expectation value of an \emph{$n$-particle local energy} $E^{\text{loc}}_n(x)$, taken jointly over a probability distribution of the particle number, denoted $P_n$, and the probability distribution of the $n$-particle wave function, denoted $|\bar{\varphi}_n(x)|^2$:
\begin{equation}\label{eq:Energy_modified_VMC}
\begin{aligned}
    E(|\Psi\rangle) = \frac{\langle \Psi | H | \Psi \rangle}{\langle \Psi | \Psi \rangle} = \mathop{\mathbb{E}}_{n\sim P_n} \mathop{\mathbb{E}}_{\textbf{x}_n \sim |\bar{\varphi}_n|^2} \Big[ E^{\text{loc}}_n(\textbf{x}_n) \Big].
\end{aligned}
\end{equation}
Explicitly, the $n$-particle local energy is $E^{\text{loc}}_n(\textbf{x}_n) = {\langle \textbf{x}_n | H | \Psi \rangle }/{\langle \textbf{x}_n | \varphi_n \rangle }$, the probability distribution of the particle number is $P_n := {\langle \varphi_n | \varphi_n \rangle}/{\langle \Psi | \Psi \rangle}$, and the probability distribution of the $n$-particle wave function is $|\bar{\varphi}_n(\textbf{x}_n)|^2 := {|\varphi_n(\textbf{x}_n)|^2}/{\langle \varphi_n | \varphi_n \rangle}$.

We estimate the above expectation value and its standard deviation by drawing samples from $P_n$ and $|\bar{\varphi}_n(x)|^2$ jointly via the \emph{Markov Chain Monte Carlo (MCMC) in Fock space} algorithm described in the Supplementary Material. This algorithm employs Metropolis-Hastings sampling with a proposal function that allows the particle number to increase or decrease, such that $P_n$ and $|\bar{\varphi}_n|^2$ are sampled simultaneously. The standard deviation of the expectation value is controllable and decays as the inverse square root of the number of samples.

We use gradient-based optimization to optimize a NQFS. As shown in the Supplementary Material, the derivative of the energy with respect to a variational parameter, say a parameter of the neural network, may also be expressed as a combination of expectation values over $P_n$ and $|\bar{\varphi}_n(x)|^2$. We again estimate these as empirical means over MCMC samples, and feed the resulting gradient estimate into the the ADAM algorithm~\cite{kingma2014adam} to minimize the energy. The systematic uncertainty of the optimized NQFS can be studied with variance extrapolation~\cite{Sorella_2001, Kashima_2001}, as explained in the Supplementary Material.

It is useful here to employ regularization and system-dependent modifications that assist with the learning procedure (see Supplementary Material for specific details). To facilitate smooth optimization over particle number, we multiply $\varphi^{\text{NQFS}}_n(\textbf{x}_n)$ by a parameterizable regularization factor that keeps the probability distribution over particle number well-behaved. Moreover, for systems with hard wall boundary conditions, we multiply $\varphi^{\text{NQFS}}_n(\textbf{x}_n)$ by a cutoff factor~\cite{sarsa2011variational, flores2008compression, sarsa2014study, laughlin2009highly} that forces it to vanish at the boundaries. Lastly, for an interaction potential that diverges as two particles approach each other $x_i \rightarrow x_j$, the eigenstates exhibit a cusp at $x_i=x_j$ that is quantified by Kato's cusp condition~\cite{kato1957eigenfunctions, foulkes2001quantum}; we account for this cusp by multiplying $\varphi^{\text{NQFS}}_n(\textbf{x}_n)$ by a Jastrow factor that satisfies Kato's cusp condition.

To highlight applicability, we apply NQFS to the Lieb-Liniger model in an inhomogenous system with hard walls, the Calogero-Sutherland model which has long-range interactions, and a regularized Klein-Gordon model that does not conserve particle number. We select these models because they have exactly solvable ground states to benchmark against (details provided in the Supplementary Material). These models are non-perturbative in that their interactions strongly influence the ground states and cannot be treated perturbatively. Beyond these models, NQFS can be straightforwardly applied to non-exactly solvable field theories, which we exemplify in the Supplementary Material.

\emph{Lieb-Liniger Model.}---
Let us illustrate the application of NQFS to the quintessential non-relativistic QFT, the Lieb-Liniger model~\cite{Lieb_1963_1, Lieb_1963_2}:
\begin{equation}\label{eq:LL_Hamiltonian}
\begin{aligned}
    H_{\text{LL}} = \ &\frac{1}{2m} \int dx \frac{d\hat{\psi}^\dag(x)}{dx} \frac{d \hat{\psi} (x)}{dx} - \mu \int dx \hat{\psi}^\dag(x) \hat{\psi} (x) \\
    & + g \int dx \hat{\psi}^\dag(x) \hat{\psi}^\dag(x) \hat{\psi}(x) \hat{\psi}(x).
\end{aligned}
\end{equation}
This model describes bosons of chemical potential $\mu$ interacting via a contact interaction of strength $g$: $V(x-y) = 2g \delta(x-y)$. Experimentally, the Lieb-Liniger model has been realized with ultracold atoms~\cite{Lieb_2003} and in optical lattices~\cite{paredes2004tonks, Kinoshita_2004}. The model is integrable and thus provides a useful benchmark for numerics; as the Hamiltonian conserves particle number, its ground state lies in a definite particle number sector, which we denote by $n_0$.

\begin{figure}[tbp]
    \includegraphics[width=0.49\textwidth]{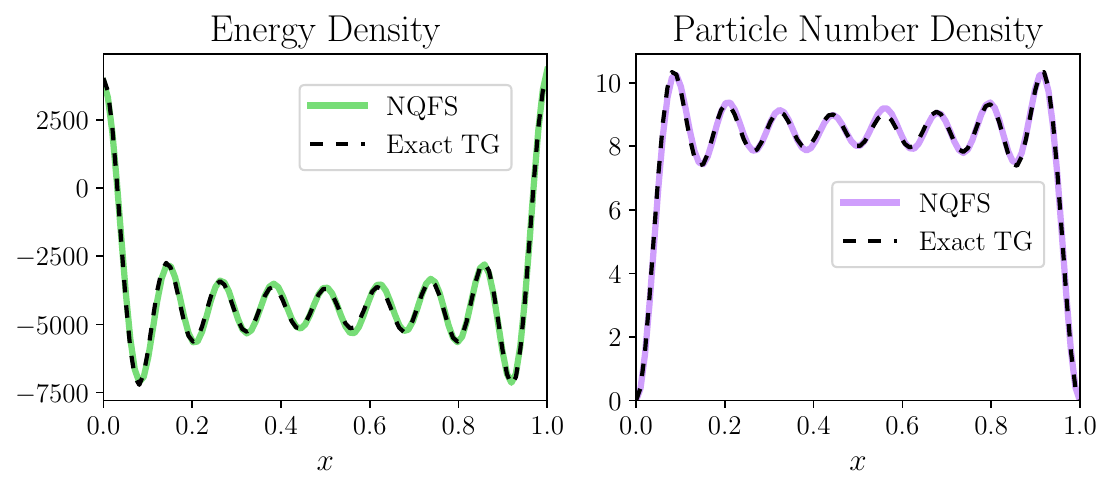}
    \caption{The energy density (\textbf{Left}) and particle number density (\textbf{Right}) of the NQFS (at $g=10^6$) and the exact Tonks-Girardeau ground state of the Lieb-Liniger model. \vspace{-16px}}
    \label{fig:LL_Plots}
\end{figure}

With our NQFS ansatz, we probe the ground state of an inhomogneous system with hard walls at $x=0$ and $x=L$, accommodating these boundary conditions with a cutoff factor as mentioned above, and accounting for the delta function potential with an appropriate Jastrow factor. We first consider the Tonks-Girardeau limit, in which $g \rightarrow \infty$ and the model can be mapped to non-interacting spinless fermions~\cite{Girardeau_1960, PhysRevLett_Tuybens}. We simulate this limit with parameters $L=1$, $m=1/2$, $\mu = (8.75 \pi)^2$, and $g = 10^6$, in which case the ground state has energy $E_0 = -4031.79$ and $n_0 = 8$ particles. After optimization, we obtain an energy $E=-4030.9 \pm 0.1$ and particle number $n = 8.00 \pm 0.01$, showcasing that NQFS are capable of finding both the ground state and the particle number sector in which it lies. We also compare the energy density and the particle number density of the NQFS with the Tonks-Girardeau solution in Fig.~\ref{fig:LL_Plots}, the clear agreement of which further justifies that the NQFS has learned the ground state correctly.

To study performance at smaller values of $g$, we set $\mu = 185$ and $g = 10$, in which case $E_0 = -954.60$ and $n_0 = 10$. Our NQFS finds $E = -952.20 \pm 0.09$ and $n = 10.00 \pm 0.03$, indicating that good performance is maintained away from the Tonks-Girardeau limit.

\emph{Calogero-Sutherland Model.}---
We further aim to study a system with long-range interactions, for which purpose we consider the Calogero-Sutherland model on a ring of length $L$~\cite{Sutherland_1971, Sutherland_1972}:
\begin{equation}\label{eq:CMS_Interaction}
\begin{aligned}
    & H_{\text{CS}} = \ \frac{1}{2m} \int dx \frac{d\hat{\psi}^\dag(x)}{dx} \frac{d \hat{\psi} (x)}{dx} - \mu \int dx \hat{\psi}^\dag(x) \hat{\psi} (x) \\
    & \ + \frac{g \pi^2}{2 L^2} \int dx dy \hat{\psi}^\dag(x) \hat{\psi}^\dag(y) \hat{\psi}(y) \hat{\psi}(x) \big[\sin(\tfrac{\pi}{L}(x-y))\big]^{-2}
\end{aligned}
\end{equation}
This model describes particles interacting via an inverse square sinusoidal potential with interaction strength $g$. The ground state energy is $ E_0 = \frac{\pi^2\lambda^2}{6mL^2}n_0(n_0^2-1) - \mu n_0$, where $\lambda = \tfrac{1}{2}\big(1+\sqrt{1+4mg}\big)$, and $n_0$ minimizes $E_0$ and is the particle number of the ground state.

\begin{figure}[tbp]
    \includegraphics[width=0.47\textwidth]{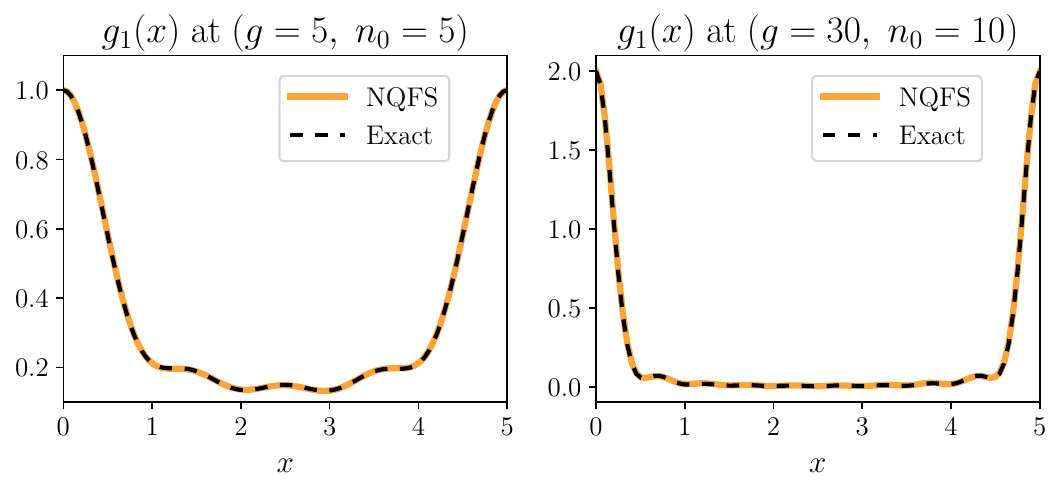}
    \caption{The NQFS and exact one-body density matrices $g_1(x)$ of the Calogero-Sutherland model at ($g=5$, $n_0=5$) (\textbf{Left}), and ($g=30$, $n_0=10$)  (\textbf{Right}). \vspace{-18px}}
    \label{fig:CS_OneBodies}
\end{figure}

We first study this model with system size $L=5$ at $g=5$ and $\mu = 3 \cdot 5^2 \cdot \frac{\pi^2\lambda^2}{6mL^2}$, which has exact ground state energy $E_0=-156.317$ and particle number $n_0=5$. Employing NQFS with an appropriate Jastrow factor, we obtain $E = -156.291 \pm 0.003$ and $n = 5.000 \pm 0.007$, indicative of great performance. We also look at the one-body density matrix $g_1(x) = n \int d^{n-1}x \varphi_n^*(x, x_2,\ldots , x_n) \varphi_n(0, x_2,\ldots , x_n)$, and compare it to the exact solution in Fig.~\ref{fig:CS_OneBodies}. 

We next consider a more strongly interacting limit with $g=30$ and $\mu = 3 \cdot 10^2 \cdot \frac{\pi^2\lambda^2}{6mL^2}$, in which case $E_0=-5132.76$ and $n_0 = 10$. NQFS yields $E= -5131.77 \pm 0.09$ and $n=10.000 \pm 0.004$, and we again illustrate the one-body density matrix in Fig.~\ref{fig:CS_OneBodies}. The agreement with the exact solution in both of these cases evidences NQFS's applicability to systems with long-range interactions.

\emph{Regularized Klein-Gordon Model.}---
The previous Hamiltonians conserve particle number and have ground states that lie in a single particle number sector. We are also interested in Hamiltonians that violate particle number conservation and whose ground states are necessarily superpositions of $n$-particle wave functions. 

For this purpose, we consider the Klein-Gordon model: $H_{\text{KG}} = \frac{1}{2}\int dx \ |\hat{\pi}(x)|^2 + |\nabla \hat{\phi}(x) |^2 + m^2|\hat{\phi}(x)|^2$, where $\hat{\phi}(x)$ and $\hat{\pi}(x)$ are canonically conjugate fields. To prevent this Hamiltonian from diverging, it is necessary to introduce a momentum cutoff $\Lambda$ and regularize $H_{\text{KG}}$; a convenient way to do so is by adding a counterterm $\frac{1}{\Lambda^2}|\nabla \hat{\pi}(x)|^2$~\cite{Stojevic_2015}. With this modification, we can introduce creation and annihilation operators by the change of variables $\hat{\phi}(x) = \frac{1}{\sqrt{2\Lambda}}\big(\hat{\psi}(x) + \hat{\psi}^\dag(x) \big)$ and $\hat{\pi}(x) = -i\sqrt{\frac{\Lambda}{2}}\big(\hat{\psi}(x) - \hat{\psi}^\dag(x)\big)$, upon which the regularized Hamiltonian is mapped to the following quadratic Hamiltonian~\cite{Stojevic_2015, Karanikolaou_2021}:
\begin{equation}
\begin{aligned}
    H_{\text{Quad}} = \ &\int dx \frac{d\hat{\psi}^\dag(x)}{dx} \frac{d \hat{\psi} (x)}{dx} + v \int dx \hat{\psi}^\dag(x) \hat{\psi} (x) \\
    & + \lambda \int dx \big(\hat{\psi}^\dag(x) \hat{\psi}^\dag(x) + \hat{\psi}(x) \hat{\psi}(x) \big),
\end{aligned}
\end{equation}
where the coefficients are $v=\frac{1}{2}(m^2 + \Lambda^2)$ and $\lambda = \frac{1}{4}(m^2 - \Lambda^2)$, and the last term violates particle number conservation. In this formulation, $|\lambda/v| \leq 1/2$ in order for the Hamiltonian to be well-defined; as $|\lambda/v| \rightarrow 1/2$, the spectrum becomes gapless. The ground state of $H_{\text{Quad}}$ can be exactly determined by moving to momentum space and performing a Bogoliubov transformation.

\begin{figure}[tbp]
    \includegraphics[width=0.49\textwidth]{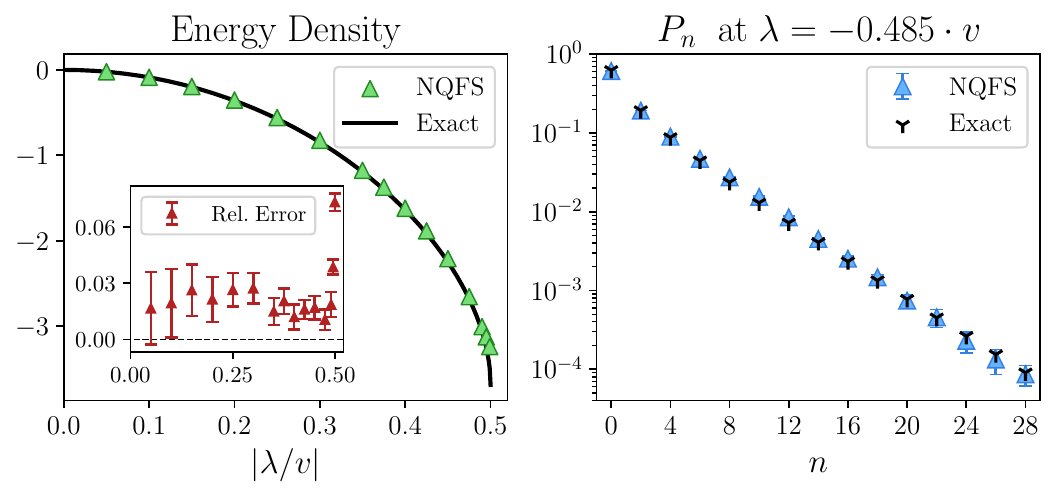}
    \caption{\textbf{Left}: The NQFS and exact energy densities of the regularized Klein-Gordon model across a range of $\lambda$, as well as the corresponding relative errors and standard deviations in the inset. \textbf{Right}: The NQFS and exact particle number distributions at $\lambda = -0.485\cdot v$.\vspace{-18px}}
    \label{fig:Quad_EnergyDensity}
\end{figure}

We study a finite system of size $L=1$ with coefficient $v=6$ over a range of $\lambda$ from $\lambda = 0$ to the critical point $\lambda = -3$. As suggested by the particle number non-conserving term in the Hamiltonian, we restrict our NQFS to $n$-particle wave functions with an even number of particles. We compare the resulting energy density of the NQFS with the exact solution in Fig.~\ref{fig:Quad_EnergyDensity}. This plot indicate good agreement with the exact solution, achieving errors around a few percent that increase as the critical point is approached. To emphasize that the NQFS has truly learned a superposition of $n$-particle wave functions, in Fig.~\ref{fig:Quad_EnergyDensity} we plot the particle number distributions $P_n$ at $\lambda = -0.485\cdot v$. We see clear agreement between the NQFS and exact distributions, evidencing that NQFS can accurately represent such a superposition.

While these results are promising, we note that NQFS struggle on systems of larger size and greater couplings $v$ and $\lambda$. As we note in the Supplementary Material, one reason for this is that the local energy of the term $ \lambda \int dx \psi^\dag(x)\psi^\dag(x) + h.c. $ imposes a sort of ``cusp condition" that is nontrivial to build into a variational ansatz. As we do not explicitly account for this condition here, the energy suffers a large variance that hinders its optimization and can result in a subpar ground state approximation; this effect becomes most pronounced near the critical point. Improving performance near the critical point provides an interesting direction for future work.

\emph{Conclusion.}---
We have developed neural-network quantum field states as a variational ansatz for quantum field theories in the continuum, and demonstrated their application to an inhomogeneous system, a system with long-range interactions, and a system that breaks particle number conservation. Our work opens the door to new variational techniques in QFT. 

Beyond the applications demonstrated here, there are ample refinements and extensions of NQFS. From an algorithmic point of view, optimization over the particle number can be hindered by $n$-particle wave functions whose magnitudes differ significantly across particle number sectors. While we mitigated this issue by incorporating a regularization factor, it would be beneficial to develop more precise algorithms for stable optimization over particle number. Furthermore, another direction for improving performance is to properly account for the cusp conditions imposed by QFT Hamiltonians, in particular those that violate particle number conservation. 

From a physics perspective, a particularly enticing idea is to extend NQFS to relativistic QFT, which would require a move to the thermodynamic limit and also a resolution to Feynman's warning about the sensitivity to high frequencies suffered by variational methods in relativistic QFT~\cite{feynman1988difficulties}. A generalization to fermionic fields would also be fascinating; to enforce antisymmetry here, one could use a Deep Sets with an antisymmetric aggregation function, such as a Slater determinant. Another interesting avenue is application to real-time dynamics of QFT, which could provide physics simulations beyond the Euclidean time formulation. Moreover, the NQFS framework can be straightforwardly adapted to $d\geq 2$ spatial dimensions by treating positions $x_i$ as $d$-dimensional vectors. This could enable simulation of higher dimensional QFTs and surpass the capabilities of traditional methods like continuous tensor networks~\cite{Jennings_2015, Tilloy2019}.

\emph{Code Availability}--- Codes to run the simulations in this Letter are available on GitHub at Ref.~\cite{NQFS_Repo}.

\emph{Acknowledgements}--- 
The authors acknowledge useful feedback and discussions from Isaac Chuang, Martin Ganahl, Bryan Clark, William Detmold, Alexander Zlokapa, Trevor McCourt, Rikab Gambhir, Aldi Maraya, Phiala Shanahan, and James Stokes. JMM acknowledges the MIT SuperCloud and Lincoln Laboratory Supercomputing Center for providing the computing resources used in this paper. JMM acknowledges support from the National Science Foundation Graduate Research Fellowship under Grant No. 2141064. DL acknowledges support from the NSF AI Institute for Artificial Intelligence and Fundamental Interactions (IAIFI). This material is based upon work supported by the U.S. Department of Energy, Office of Science, National Quantum Information Science Research Centers, Co-design Center for Quantum Advantage (C2QA) under contract number DE-SC0012704.

\bibliography{References}
\bibliographystyle{apsrev4-1}

\appendix

\clearpage

\onecolumngrid
\begin{center}
	\noindent\textbf{Supplementary Material}
	\bigskip
		
	\noindent\textbf{\large{}}
\end{center}

\twocolumngrid

\section{Non-relativistic Field Theory}
As the focal point of our work is non-relativistic QFT of bosons, let us briefly review this topic. For a more thorough and pedagogical introduction to non-relativistic QFT, see Ref.~\cite{altland2010condensed}. 

Non-relativistic QFT is described by second quantization. In the bosonic formulation of second quantization, particles are created and annihilated by operators $\hat{\psi}^\dag(x)$ and $\hat{\psi}(x)$, respectively, which in 1D obey the bosonic commutation relation $[\hat{\psi}(x), \hat{\psi}^\dag(x')] = \delta(x-x')$. Quantum field states live in a Fock space that is a direct sum over all symmetrized $n$-particle Hilbert spaces, such that an arbitrary state $|\Psi \rangle$ in this space may be expressed as a superposition of permutation invariant $n$-particle wave functions $\varphi_n(\textbf{x}_n)$:
\begin{equation}\label{eq:QFT_State_FockSpace}
\begin{aligned}
    |\Psi\rangle = \bigoplus_{n=0}^\infty |\varphi_n \rangle = \bigoplus_{n=0}^\infty \int d^nx \ \varphi_n(\textbf{x}_n) |\textbf{x}_n\rangle,
\end{aligned}
\end{equation}
where we use the shorthand $\textbf{x}_n = (x_1, x_2, \dots, x_n)$, and employ the $n$-particle basis
\begin{equation}
\begin{aligned}
    |\textbf{x}_n \rangle &= |x_1, x_2, \ldots, x_n \rangle \\
    &= \frac{1}{\sqrt{n!}} \hat{\psi}(x_1)^\dag \hat{\psi}(x_2)^\dag \dots \hat{\psi}(x_n)^\dag |\Omega\rangle,
\end{aligned}
\end{equation}
where $|\Omega \rangle$ is the vacuum state. The wave functions $\varphi_n(\textbf{x}_n)$ are unnormalized in this formulation.  

By construction, $|\textbf{x}_n\rangle$ is symmetric under exchange of any two particles. The field creation and annihilation operators act on $|\textbf{x}_n \rangle$ as:
\begin{equation}
\begin{aligned}
    &\hat{\psi}^\dag(x) |\textbf{x}_n \rangle = \sqrt{n+1} |x_1, x_2..., x_n, x\rangle \\
    & \hat{\psi} (x) |\textbf{x}_n \rangle = \frac{1}{\sqrt{n}} \sum_{i=1}^n \delta(x_i=x) |x_1,..., x_{i-1}, x_{i+1},..., x_n \rangle.
\end{aligned}
\end{equation}
The $|\textbf{x}_n \rangle$ basis is symmetrically orthonormalized as
\begin{equation}
    \langle \textbf{x}'_{n'} | \textbf{x}_n \rangle = \delta_{n, n'} \frac{1}{n!} \sum_{\textbf{x}''_n \in \text{Sym}(\textbf{x}'_n)} \delta^n(\textbf{x}''_n - \textbf{x}_n) , 
\end{equation}
where the sum runs over all permutations ${\textbf{x}}''_n \in \text{Sym}(\textbf{x}'_n)$ of $\textbf{x}'_n$, such that a permutation invariant $n$-particle wave function $\varphi_n (\textbf{x}_n)$ may be extracted as
\begin{equation}
    \langle \textbf{x}_n | \Psi \rangle = \varphi_n (\textbf{x}_n). 
\end{equation}

The Hamiltonian of a non-relativistic QFT is a function of the creation and annihilation operators, typically taking the form of an integral over space of a Hamiltonian density. The prototypical such Hamiltonian contains a kinetic energy term, an onsite external potential $V(x)$, a chemical potential $\mu$, and a two-body interaction potential $W(x-y)$:
\begin{equation}\label{eq:QFT_Hamiltonian1}
\begin{aligned}
    H = \ &\frac{1}{2m} \int dx \frac{d\hat{\psi}^\dag(x)}{dx} \frac{d \hat{\psi} (x)}{dx} \\
    & + \int dx \big( V(x)-\mu \big) \hat{\psi}^\dag(x) \hat{\psi} (x) \\
    & + \frac{1}{2} \int dx dy W(x-y) \hat{\psi}^\dag(x) \hat{\psi}^\dag(y) \hat{\psi}(y) \hat{\psi}(x).
\end{aligned}
\end{equation}
This can be viewed as a quantum mechanical Hamiltonian in the grand canonical ensemble, where the number of particles is not fixed, but instead dictated by the chemical potential. We note that this prototypical Hamiltonian conserves particle number, and thus its ground state lies in a single $n$-particle sector, i.e. $|\Psi_0\rangle = \int d^nx \ \varphi_n(\textbf{x}_n) |\textbf{x}_n \rangle$ for some particle number $n$.

Also relevant are Hamiltonians that do not conserve particle number. For instance, in the main text, we studied the Hamiltonian
\begin{equation}
\begin{aligned}
    H = \ &\int dx \frac{d\hat{\psi}^\dag(x)}{dx} \frac{d \hat{\psi} (x)}{dx} + v \int dx \hat{\psi}^\dag(x) \hat{\psi} (x) \\
    & + \lambda \int dx \big(\hat{\psi}^\dag(x) \hat{\psi}^\dag(x) + \hat{\psi}(x) \hat{\psi}(x) \big).
\end{aligned}
\end{equation}
The last term violates particle number conservation, and consequently its ground state is necessarily a superposition of $n$-particle wave functions.

\section{Continuous Matrix Product States}
Approximating the ground state of non-relativistic field theory is a nontrivial problem, as it necessitates optimization over a state in Fock space. For a Hamiltonian that conserves particle number, one could use a standard variational approach from quantum mechanics to iteratively approximate the ground states over a range of $n$-particle sectors, and select the state that achieves the lowest energy. While certainly functional, this approach is slow, and more severely it is inapplicable to Hamiltonians that violate particle number conservation. A swifter and more applicable approach is to directly parameterize a state in Fock space and optimize over all of its $n$-particle wave functions simultaneously.

Prior research in this direction has led to the development of the continuous matrix product state (cMPS)~\cite{Verstraete2010, Haegeman_2013_calculus}, a powerful variational ansatz for 1D field theories. A cMPS is obtained by taking the continuum limit of a matrix product state (MPS) on a lattice -- that is, taking the limit as the lattice spacing goes to $0$ while keeping the system size, $L$, fixed. Performing this limit carefully to avoid divergences, one obtains the following cMPS on a periodic system (see Ref.~\cite{Haegeman_2013_calculus} for modifications to other boundary conditions):
\begin{equation}
    |\Psi^{\text{cMPS}}\rangle = \text{Tr}_{\text{aux}} \Big[ \mathcal{P} e^{\int_0^L dx (Q(x) \otimes I + R(x) \otimes \hat{\psi}(x)^\dag)} \Big] |\Omega \rangle.
\end{equation}
In this expression, $Q(x)$ and $R(x)$ are dimension $D\times D$ matrix-valued functions that parameterize the cMPS, and $D$ is known as the bond dimension. The operation $\text{Tr}_{\text{aux}}$ denotes a trace over the $D\times D$-dimensional auxiliary space in which $Q(x)$ and $R(x)$ exist, and $\mathcal{P}$ denotes the path ordering of the exponential. 

Fundamentally, a cMPS represents a state in Fock space with unnormalized $n$-particle wave functions 
\begin{equation}
\begin{aligned} 
   \varphi_n^{\text{cMPS}}(\textbf{x}_n) = \sqrt{n!} \cdot \text{Tr}\Big( u_Q(0,x_1&)R( x_1) u_Q(x_1,x_2)R(x_2)\\ 
   & \ \ \dots R(x_N)u_Q(x_N,L) \Big), 
\end{aligned}
\end{equation}
where $u_Q(x_i,x_j) = \mathcal{P}e^{\int_{x_i}^{x_j}Q(x) dx}$. Analogous to their discrete cousins, cMPSs are dense in Fock space and exhibit area law scaling of their entanglement entropy, theoretically rendering them a good candidate for ground state approximation of field theories~\cite{tilloynotes}. Expectation values of a cMPS state may be expressed in terms of $Q(x)$, $R(x)$, and derivatives thereof; accordingly, $Q(x)$ and $R(x)$ are optimized over to pinpoint the ground state.

Despite the promise of cMPS, in practice they face limitations due to troublesome optimization and a restricted range of applications. While efficient gradient-based optimization algorithms have been developed for cMPS in translation-invariant systems~\cite{Ganahl_2017}, in inhomogenous settings, optimization of cMPS via the time dependent variational principle requires the solution to a non-linear matrix-valued partial differential equation whose integration is prone to numerical instabilities~\cite{Haegeman_2017, PhysRevLett_Tuybens}. Refs.~\cite{Ganahl_2017_Continuous, Ganahl2018PhRvB, PhysRevLett_Tuybens} have proposed algorithms to better adapt cMPS to inhomogenous systems, but ultimately rely on lattice discretizations and/or interpolations, rather than work directly in the continuum. Moreover, cMPS cannot be straightforwardly applied to systems with long-range interactions, and have most often been restricted to systems with contact interactions, i.e. $W(x-y) \propto \delta(x-y)$. To the best of our knowledge, only Ref.~\cite{Rincon_2015_Lieb} applies cMPS to a translation invariant system with an exponentially decaying interaction potential, and very recently, Ref.~\cite{Lukin_2022} developed an approximate method for applying cMPS to long-range interactions, which employs the discretization scheme of Ref.~\cite{PhysRevLett_Tuybens} and approximates the interaction potential as a sum of exponentials. Furthermore, cMPS can be extended to higher higher spatial dimensions through the formalism of continuous tensor networks~\cite{Tilloy2019, Jennings_2015}; these however are susceptible to UV divergences and have not yet been successfully applied to field theories away from the non-interacting limit~\cite{Karanikolaou_2021, Shachar_2022}.

\section{Neural-Network Quantum States}
Spawned by recent advances in machine learning, variational ansatzes based on neural networks have emerged under the name \emph{neural-network quantum states} (NQS). These states parameterize a wave function by neural networks, which are subsequently optimized over to pinpoint a state of interest. Typically, the energy of an NQS is estimated and minimized with variational Monte Carlo~\cite{Vicentini_2022}. Among the variety of neural network architectures successfully employed in NQS include restricted Boltzmann machines~\cite{Carleo2017, melko2019restricted}, feedforward neural networks~\cite{Choo_2018}, and recurrent neural networks~\cite{hibat2020recurrent}. 

In support of these ansatzes, Refs.~\cite{Deng_2017, Glasser_2018, Levine_2019} have shown that NQS can efficiently represent highly entangled quantum many-body states, and Refs.~\cite{Chen_2018, sharir2021neural} that they are capable of performing accurate tensor contractions with little overhead. Consequently, NQS have been met with success in spin systems~\cite{Carleo2017, melko2019restricted, Choo_2018, Saito_2018, Sharir_2020, hibat2020recurrent}, open systems~\cite{Yoshioka_2019, Vicentini_2019, Nagy_2019, Hartmann_2019}, atomic nuclei~\cite{Lovato_2022, Gnech_2021, Adams_2021}, quantum chemistry~\cite{Hermann_2020, Pfau_2020}, and continuous space systems~\cite{Pescia_2022,wilson2022wave}.

\section{Variational Monte Carlo}
Variational Monte Carlo (VMC) is a statistical method for estimating the energy of a quantum mechanical state. Coupled with the variational principle, VMC allows one to approximate the ground state of a system. VMC is predicated on the relation that the energy of a state $|\psi\rangle$ may be expressed as an expectation value of a \emph{local energy} $E^{\text{loc}}(\textbf{x})$ over the probability distribution of the position space wave function:
\begin{equation}
\begin{aligned}
    & E(|\psi\rangle) = \frac{\langle \psi | H | \psi \rangle }{\langle \psi | \psi \rangle} = \mathop{\mathbb{E}}_{\textbf{x} \sim |\bar{\psi}(x)|^2} \big[ E^{\text{loc}}(\textbf{x}) \big], \\ 
    & |\bar{\psi}(\textbf{x})|^2 := \frac{|\psi(\textbf{x})|^2}{\int d\textbf{x}' |\psi(\textbf{x}')|^2},
\end{aligned}
\end{equation}
where $E^{\text{loc}}(\textbf{x}) = \frac{\langle \textbf{x} | H | \psi\rangle }{\langle \textbf{x} | \psi\rangle }$ is the local energy of configuration $\textbf{x}$. In practice, one often uses Markov Chain Monte Carlo (MCMC) sampling to generate samples $\{x^{(j)}\}_{j=1}^N$ from $|\bar{\psi}(\textbf{x})|^2$ and estimate this expectation value as 
\begin{equation}\label{eq:VMC_Energy}
    E(|\psi\rangle) = \mathop{\mathbb{E}}_{\textbf{x}\sim |\bar{\psi}(x)|^2} \big[ E^{\text{loc}}(\textbf{x}) \big] \approx \frac{1}{N} \sum_{j=1}^N E^{\text{loc}}(\textbf{x}^{(j)}).
\end{equation}

Similarly, the derivative of the energy with respect to a variational parameter $\alpha$ may be expressed as
\begin{equation}
\begin{aligned}
    \frac{\partial}{\partial \alpha} E(|\psi\rangle) = 
    2 \cdot \text{Re} \Bigg(\frac{\langle  \psi |H| \partial_\alpha \psi \rangle}{\langle \psi | \psi \rangle} - E(\psi)\frac{\langle \psi| \partial_\alpha \psi \rangle}{\langle \psi | \psi \rangle}\Bigg),
\end{aligned}
\end{equation}
where the terms comprising this expression can be evaluated as expectation values:
\begin{equation}\label{eq:VMC_Energy_Derivative}
\begin{aligned}
    &\frac{\langle  \psi |H| \partial_\alpha \psi \rangle}{\langle \psi | \psi \rangle} =  \mathop{\mathbb{E}}_{\textbf{x}\sim |\bar{\psi}(x)|^2} \left[ \frac{\langle \textbf{x} | H | \partial_\alpha \psi\rangle }{\langle \textbf{x} | \psi\rangle } \right], \\
    &\frac{\langle \psi| \partial_\alpha \psi \rangle}{\langle \psi | \psi \rangle} = \mathop{\mathbb{E}}_{\textbf{x}\sim |\bar{\psi}(x)|^2} \left[ \frac{\langle \textbf{x} | \partial_\alpha \psi\rangle }{\langle \textbf{x} | \psi\rangle } \right] .
\end{aligned}
\end{equation}
These quantities can also be estimated as empirical means of MCMC samples to obtain an estimate of the gradient w.r.t. the variational parameters. One can then feed this approximate gradient into a gradient-based optimization algorithm to minimize the energy and approximate the ground state.

\section{Deep Sets}
In the main text, we constructed our NQFS ansatz from a neural network architecture known as \emph{Deep Sets}. Initially pioneered in Ref.~\cite{2017_Zaheer}, the Deep Sets architecture seeks to model a function of a set $X$. As such a set is not ordered, this function is necessarily permutation invariant. Ref.~\cite{2017_Zaheer} proves that any such function $f(X)$ can be decomposed as
\begin{equation}\label{eq:DeepSetsDecomp}
    f(X) = \rho\Big(\sum_{x \in X} \phi(x)\Big),
\end{equation}
where $\rho$ and $\phi$ are constituent functions that depend on $f$. The function $\phi$ maps the inputs into a feature space of possibly higher dimension, the sum aggregates the embedded inputs in a permutation invariant manner, and $\rho$ adds the correlations necessary to output the function of interest. 

For example, consider a set $\{ x_i \}_{i=1}^n$. The product of the set $\prod_{i=1}^n x_i$ can be computed with constituent functions $\phi(x) = \ln(x)$ and $\rho(x) = e^{x}$. Similarly, if $x_i$ represents a displacement along the $i^{\mathrm{th}}$ dimension, the total Euclidean distance $\sqrt{\sum_{i=1}^n x_i^2}$ can be computed with constituent functions $\phi(x) = x^2$  and $\rho(x) = \sqrt{x}$.

In machine learning, this decomposition has been used to learn permutation invariant functions by parameterizing $\rho$ and $\phi$ as deep neural networks, and optimizing over them to pinpoint a function of interest~\cite{2017_Zaheer}. Deep Sets have also recently been used in NQS contexts to model wave functions that are permutation invariant, such as ground states whose of atomic nuclei~\cite{Lovato_2022, Gnech_2021, Adams_2021} and bosons at fixed particle number~\cite{Pescia_2022}.

Another key property of Deep Sets is that they are \emph{variadic}: they can accept an arbitrary number of arguments. To see this, note that an arbitrary number of inputs may be included in the sum in Eq.~\eqref{eq:DeepSetsDecomp}. This property was noted in Ref.~\cite{Pescia_2022} and used to approximate the ground state of a system with few particles before proceeding to a system with more particles. In the main text we leveraged this variadic property to develop a neural network ansatz for continuum QFT, where the number of particles is arbitrary.

\section{Neural-Network Quantum Field States (NQFS)}
Let us lay out our construction of a NQFS. Using the Fock space representation of a quantum field state (Eq.~\eqref{eq:QFT_State_FockSpace}), we choose to parameterize each $n$-particle wave function as a product of two Deep Sets -- one that accounts for particle positions $\{x_i\}_{i=1}^n$, and the other for particle separations $\{x_i - x_j\}_{i<j}$:
\begin{equation}\label{eq:NQFS_definition2}
\begin{aligned}
    &|\Psi^{\text{NQFS}}\rangle = \bigoplus_{n=0}^\infty \int d^nx \ \varphi_n^\text{NQFS}(\textbf{x}_n) |\textbf{x}_n\rangle, \\
    &\varphi_n^\text{NQFS}(\textbf{x}_n) = \frac{1}{L^{n/2}} \cdot f_1\big( \{x_i\}_{i=1}^n \big) \cdot f_2\big( \{x_i-x_j\}_{i < j} \big).
\end{aligned}
\end{equation}
As we noted in the main text, the factor of $\tfrac{1}{L^{n/2}}$ is included for dimensional consistency, and the constituent functions of $f_1$ and $f_2$ ($\rho_1, \phi_1$ and $\rho_2, \phi_2$, respectively) are parameterized as feedforward neural networks. We note that a phase could be incorporated via a multiplicative factor $e^{i f_3( \{x_i\} )}$ where $f_3$ is an additional Deep Sets, but this term can be set to 1 for the applications of the main text. We also note that the computational complexity of this ansatz scales polynomially as $O(n^2)$ because we consider all pairs of $n$ particles, and hence there is no exponential barrier in going to a larger particle number.

In the main text, we employed feature embeddings before feeding inputs into the neural networks. In a closed system with hard walls at $x=0$ and $x=L$, we normalize inputs by using the embedding $x_i \mapsto (\frac{x_i}{L}, \ 1-\frac{x_i}{L})$ which measures the distance to both edges of the system, and $(x_i-x_j) \mapsto \big(\frac{x_i-x_j}{L}\big)^2$, which is even to maintain permutation invariance. In a periodic system, we use $x_i \mapsto \big(\sin(\frac{2\pi }{L} x_i ) ,\ \cos(\frac{2\pi }{L} x_i)\big)$, and $(x_i-x_j) \mapsto \cos(\frac{2\pi }{L}(x_i-x_j))$. An alternative embedding for a periodic systems, as pursued in Ref.~\cite{Pescia_2022}, is the set of the first $K$ Fourier modes: $x_i \mapsto \big(\sin(\frac{2\pi k}{L} x_i ) ,\ \cos(\frac{2\pi k}{L} x_i)\big)_{k=1}^K$, and analogously for $x_i-x_j$.

We emphasize that the variadic property of Deep Sets is crucial to our ansatz as it enables us to parameterize the infinite collection of $n$-particle wave functions that each accept a different number of arguments, i.e. a parameterization for the infinite collection $\{\varphi_0, \ \varphi_1(x_1), \ \varphi_2(x_1, x_2), \ \varphi_3(x_1, x_2, x_3), ...\}$ rather than just a single function. To the best of our knowledge, no prior work has used this property of Deep Sets to simultaneously parameterize a family of wave functions that each accept a different number of arguments. In addition, implicit in the NQFS ansatz is the assumption that the $n$-particle wave functions comprising the ground state share a similar structure that is learned through optimization.

By appealing to the universality of Deep Sets for modelling permutation invariant functions~\cite{2017_Zaheer}, we expect that this ansatz can accurately approximate complex bosonic states. These was verified in the experiments in the main text, where we applied NQFS to models with strong interactions that cannot be treated perturbatively. However, while neural networks are universal approximators, it is well known that determining a generic ground state is a QMA-hard problem (difficult even for quantum computers), which indicates that it would be hard to exactly represent an arbitrary ground state of a field theory.

Lastly, we note that NQFS is a general framework, and that our Deep Sets construction provides one realization of it. This implies that there is room for improvement in the architecture design. For example, currently there is no symmetry (such as gauge symmetry) encoded in Deep Sets (aside from permutation invariance), which could make it hard to model ground states of field theories with symmetry. Incorporating such a symmetry could make it easier for NQFS to represent these ground states and would make for an interesting future study.

\section{Variational Monte Carlo in Fock Space}
Inspired by the success of VMC, we would like to develop an analogous statistical method for estimating and minimizing the energy of a quantum field state in Fock space. We term this approach \emph{Variational Monte Carlo in Fock Space}, and now present its construction.

The crux of this method is the relation that the energy of a quantum field state, $|\Psi\rangle$, in the Fock space representation of Eq.~\eqref{eq:QFT_State_FockSpace} can be expressed as an expectation value of an \emph{$n$-particle local energy}:
\begin{equation}
    E(| \Psi \rangle ) = \frac{\langle \Psi | H | \Psi \rangle }{\langle \Psi | \Psi\rangle} = \mathop{\mathbb{E}}_{n \sim P_n} \mathop{\mathbb{E}}_{\textbf{x}_n \sim |\bar{\varphi}_n|^2} \big[ E^{\text{loc}}_n(\textbf{x}_n) \big].
\end{equation}
In this expression, the $n$-particle local energy of the configuration $\textbf{x}_n$ is
\begin{equation}
    E^{\text{loc}}_n(\textbf{x}_n) = \frac{\langle \textbf{x}_n | H | \Psi \rangle }{\langle \textbf{x}_n | \varphi_n \rangle },
\end{equation}
the probability distribution of the particle number is
\begin{equation}
    P_n := \frac{\langle \varphi_n | \varphi_n \rangle}{\langle \Psi | \Psi \rangle}  = \frac{\int d^nx |\varphi_n(\textbf{x}_n)|^2}{\sum_{m=0}^\infty \int d^mx' |\varphi_m(\textbf{x}'_m)|^2}
\end{equation}
and the probability distribution of the $n$-particle wave function is
\begin{equation}
    |\bar{\varphi}_n (\textbf{x}_n)|^2 := \frac{|\varphi_n(\textbf{x}_n)|^2}{\int d^n x' |\varphi_n(\textbf{x}'_n)|^2}.
\end{equation}
This expression can be seen as the appropriate generalization of the VMC expression in Eq.~\eqref{eq:VMC_Energy}, with the modification that an expectation value be taken over the particle number distribution in addition to an expectation value over the distribution of the $n$-particle wave function. 

As in traditional VMC, we can estimate the energy as an empirical mean of samples $\{ \{ \textbf{x}_n^{(i_n)}\}_{i_n=1}^{N_n} \}_{n}$ drawn jointly from $P_n$ and $|\bar{\varphi}_n (\textbf{x}_n)|^2$:
\begin{equation}
\begin{aligned}
    E(| \Psi \rangle) & = 
    \mathop{\mathbb{E}}_{n \sim P_n} \mathop{\mathbb{E}}_{\textbf{x}_n \sim |\bar{\varphi}_n|^2} \big[ E^{\text{loc}}_n(\textbf{x}_n) \big] \\
    & \approx \frac{1}{\sum_n N_n} \sum_{n} \sum_{i_n=1}^{N_n} E^{\text{loc}}_n(\textbf{x}_n^{(i_n)}),
\end{aligned}
\end{equation}
Samples drawn jointly from $P_n$ and $|\bar{\varphi}_n (\textbf{x}_n)|^2$ may be obtained with an appropriate MCMC sampling algorithm, that crucially allows the particle number to change. We present such an MCMC algorithm, titled \emph{Markov Chain Monte Carlo Sampling in Fock Space}, later in the Supplementary Material.

Analogous to Eq.~\eqref{eq:VMC_Energy_Derivative}, the derivative of the energy with respect to a variational parameter $\alpha$ (say, a parameter of the neural networks comprising $|\Psi^{\text{NQFS}}\rangle$) may be expressed as 
\begin{equation}
\begin{aligned}
    \frac{\partial}{\partial \alpha} E(|\Psi \rangle) = 
    2 \cdot \text{Re} \Bigg(\frac{\langle  \Psi |H| \partial_\alpha \Psi \rangle}{\langle \Psi | \Psi \rangle} - E(\Psi)\frac{\langle \Psi| \partial_\alpha \Psi \rangle}{\langle \Psi | \Psi \rangle}\Bigg),
\end{aligned}
\end{equation}
where the terms constituting this expression can also be evaluated as expectation values over $P_n$ and $|\bar{\varphi}_n(x)|^2$:
\begin{equation}
\begin{aligned}
    &\frac{\langle  \Psi |H| \partial_\alpha \Psi \rangle}{\langle \Psi | \Psi \rangle} =  \mathop{\mathbb{E}}_{n\sim P_n} \mathop{\mathbb{E}}_{\textbf{x}_n \sim |\bar{\varphi}_n|^2} \left[ \frac{\langle \textbf{x}_n | H | \partial_\alpha \Psi\rangle }{\langle \textbf{x}_n | \Psi\rangle } \right], \\
    &\frac{\langle \Psi| \partial_\alpha \Psi \rangle}{\langle \Psi | \Psi \rangle} = \mathop{\mathbb{E}}_{n\sim P_n} \mathop{\mathbb{E}}_{\textbf{x}_n \sim |\bar{\varphi}_n|^2} \left[ \frac{\langle \textbf{x}_n | \partial_\alpha \Psi\rangle }{\langle \textbf{x}_n | \Psi\rangle } \right] .
\end{aligned}
\end{equation}
We can again estimate these expectation values as empirical means over MCMC samples drawn from $P_n$ and $|\bar{\varphi}_n|^2$, and feed the results into a gradient-based optimization algorithm to minimize the energy and approximate the ground state.

\subsubsection{Expressions for Energy}\label{sec:VMC_FockSpace_Energy}
To better elucidate VMC in Fock space, let us explicitly illustrate the VMC expressions for the energy. We first look at the kinetic energy, potential energy, and interaction energy, using the prototypical Hamiltonian of Eq.~(\ref{eq:QFT_Hamiltonian1}).

The expectation value of the kinetic energy may be expressed as
\begin{equation}\label{eq:KE}
\begin{aligned}
    &\frac{1}{\langle \Psi | \Psi \rangle} \Big\langle \Psi\Big| \frac{1}{2m} \int dx \frac{d\hat{\psi}^\dag(x)}{dx} \frac{d \hat{\psi} (x)}{dx} \Big| \Psi \Big\rangle \qquad \qquad \qquad \qquad \\
        &\qquad = \frac{1}{\langle \Psi | \Psi \rangle} \sum_{n=0}^\infty \int d^nx \tfrac{1}{2m} |\nabla \varphi_n(\textbf{x}_n)|^2 \\
        &\qquad =\mathop{\mathbb{E}}_{n\sim P_n} \mathop{\mathbb{E}}_{\textbf{x}_n \sim |\bar{\varphi}_n|^2} \Big[\tfrac{1}{2m} \big|\nabla \ln \varphi_n(\textbf{x}_n) \big|^2 \Big].
\end{aligned}
\end{equation}
While this expression correctly estimates the kinetic energy, in practice it will incur a large variance, even as $|\Psi\rangle$ approaches the ground state. This occurs because the corresponding local energy, 
\begin{equation}
\begin{aligned}
    &E_n^{\text{loc}}(\textbf{x}_n) = \\
    & \quad \tfrac{1}{2m}\left(\tfrac{\nabla \varphi_n(\textbf{x}_n)}{\varphi_n(\textbf{x}_n)}\right)^2 + \sum_{i} (V(x_i) -\mu) + \sum_{i<j} W(x_i-x_j),
\end{aligned}
\end{equation}
necessarily depends on the configuration $\textbf{x}_n$ even if $\varphi_n(\textbf{x}_n)$ is exactly the ground state. This induces a necessarily nonzero variance in the expectation value of the local energy.

To mitigate this issue, one may perform integration by parts and instead use the Laplacian expression for the kinetic energy. The resulting local energy is 
\begin{equation}
\begin{aligned}
    &E_n^{\text{loc}}(\textbf{x}_n) = \\
    &\quad \tfrac{1}{\varphi_n(\textbf{x}_n)} \Big( \tfrac{-1}{2m}\nabla^2 + \sum_{i} (V(x_i)-\mu) \\
    & \qquad \qquad \qquad \quad \quad + \sum_{i<j} W(x_i-x_j) \Big) \varphi_n(\textbf{x}_n).
\end{aligned}
\end{equation} 
As $\varphi_n(\textbf{x}_n)$ approaches the ground state, this expression approaches the ground state energy and thus becomes independent of the configuration $\textbf{x}_n$. In practice then, this expression for the kinetic energy suffers less variance than the previous formula, and for this reason we estimate the kinetic energy as 
\begin{equation}\label{eq:KE2}
\begin{aligned}
    &\frac{1}{\langle \Psi | \Psi \rangle} \Big\langle \Psi\Big| \frac{1}{2m} \int dx \frac{d\hat{\psi}^\dag(x)}{dx} \frac{d \hat{\psi} (x)}{dx} \Big| \Psi \Big\rangle \qquad \qquad \qquad \qquad  \\
        &\quad = \frac{1}{\langle \Psi | \Psi \rangle} \sum_{n=0}^\infty \int d^nx \tfrac{-1}{2m} \varphi_n^*(\textbf{x}_n) \nabla^2 \varphi_n(\textbf{x}_n) \\
        &\quad = \mathop{\mathbb{E}}_{n\sim P_n} \mathop{\mathbb{E}}_{\textbf{x}_n \sim |\bar{\varphi}_n|^2} \Big[\tfrac{-1}{2m} \Big(\nabla^2 \ln \varphi_n(\textbf{x}_n) \\ 
        & \quad \qquad \qquad \qquad \qquad \qquad + \big(\nabla \ln \varphi_n(\textbf{x}_n) \big)^2 \Big) \Big].
\end{aligned}
\end{equation}

The external and chemical potential terms may be similarly evaluated as
\begin{equation}\label{eq:1_body}
\begin{aligned}
    &\frac{1}{\langle \Psi | \Psi \rangle} \Big\langle \Psi \Big| \int dx \ (V(x)-\mu) \hat{\psi}^\dag(x) \hat{\psi} (x) \Big| \Psi \Big\rangle  \\
        &\quad = \frac{1}{\langle \Psi | \Psi \rangle} \sum_{n=0}^\infty \int d^nx \ |\varphi_n(\textbf{x}_n)|^2 \Big(\textstyle\sum_{i=1}^n (V(x_i) - \mu)\Big) \\ 
        &\quad =\mathop{\mathbb{E}}_{n\sim P_n} \mathop{\mathbb{E}}_{\textbf{x}_n \sim |\bar{\varphi}_n|^2} \Big[ \textstyle\sum_{i=1}^n (V(x_i)-\mu) \Big],
\end{aligned}
\end{equation}
and the two-body interaction potential term as
\begin{equation}\label{eq:2_body}
\begin{aligned}
    & \frac{1}{\langle \Psi | \Psi \rangle} \Big\langle \Psi \Big| \frac{1}{2} \int dx dy \ W(x-y) \hat{\psi}^\dag(x) \hat{\psi}^\dag(y) \hat{\psi}(y) \hat{\psi}(x) \Big| \Psi \Big\rangle \\
        &\quad = \frac{1}{\langle \Psi | \Psi \rangle} \sum_{n=0}^\infty \int d^nx \ |\varphi_n(\textbf{x}_n)|^2 \Big(\tfrac{1}{2} \textstyle\sum_{i\neq j} W(x_i - x_j)\Big) \\ 
        &\quad =\mathop{\mathbb{E}}_{n\sim P_n} \mathop{\mathbb{E}}_{\textbf{x}_n \sim |\bar{\varphi}_n|^2} \Big[ \textstyle\sum_{i<j} W(x_i - x_j) \Big].
\end{aligned}
\end{equation}

Next, for the regularized Klein-Gordon studied in the main text, we may express the expectation value of the term $\lambda \int dx \big( \hat{\psi}^\dag(x)\hat{\psi}^\dag(x) + \hat{\psi}(x) \hat{\psi}(x) \big)$ as 
\begin{equation}\label{eq:lam}
\begin{aligned}
    &\frac{1}{\langle \Psi | \Psi \rangle} \Big\langle \Psi\Big| \lambda \int dx' \Big( \hat{\psi}^\dag(x')\hat{\psi}^\dag(x') + \hat{\psi}(x') \hat{\psi}(x') \Big) \Big| \Psi \Big\rangle \qquad \qquad \qquad \qquad \\
        &\quad = \frac{\lambda}{\langle \Psi | \Psi \rangle} \sum_{n=0}^\infty \sqrt{(n+1)(n+2)} \int d^nx dx'  \\
        &\quad \qquad \qquad \qquad \qquad \times \varphi_n(\textbf{x}_n) \varphi_{n+2}(\textbf{x}_n, x', x')  + \text{h.c.}\\
        &\quad = \mathop{\mathbb{E}}_{n\sim P_n} \mathop{\mathbb{E}}_{\textbf{x}_n \sim |\bar{\varphi}_n|^2} \Bigg[ 2\lambda \sqrt{(n+1)(n+2)} \\
        &\qquad \qquad \qquad \qquad \qquad \quad \times \text{Re} \Big( \frac{\int dx' \varphi_{n+2}(\textbf{x}_{n}, x', x')}{\varphi_{n}(\textbf{x}_n)} \Big) \Bigg],
\end{aligned}
\end{equation}
where the integral over $x'$ may be evaluated with numerical integration, such as the trapezoid rule.

\subsubsection{Expressions for the Gradient of Energy}
With an eye towards energy minimization, let us also detail expressions for the derivative of the energy with respect to a variational parameter $\alpha$. As we explained above, this may be evaluated as
\begin{equation}
\begin{aligned}
    \frac{\partial}{\partial \alpha} E(| \Psi \rangle ) = 
    2 \cdot \text{Re} \Bigg(\frac{\langle  \Psi |H| \partial_\alpha \Psi \rangle}{\langle \Psi | \Psi \rangle} - E(\Psi)\frac{\langle \Psi| \partial_\alpha \Psi \rangle}{\langle \Psi | \Psi \rangle}\Bigg).
\end{aligned}
\end{equation}
Each term in this equation may be expressed as an expectation value over $P_n$ and $|\bar{\varphi}_n|^2$. 

The the evaluation of $E(|\Psi\rangle)$ was already presented above. The next simplest term to evaluate is 
\begin{equation}
\begin{aligned}
    &\frac{1}{\langle \Psi | \Psi \rangle} \langle \Psi| \partial_\alpha \Psi \rangle = \qquad \qquad \qquad \qquad \qquad \qquad \\
    &\qquad = \frac{1}{\langle \Psi | \Psi \rangle} \sum_{n=0}^\infty \int d^n x \  \varphi_n^*(\textbf{x}_n) (\partial_\alpha \varphi_n(\textbf{x}_n)) \\
    &\qquad = \mathop{\mathbb{E}}_{n\sim P_n} \mathop{\mathbb{E}}_{\textbf{x}_n \sim |\bar{\varphi}_n|^2} \Big[ \partial_\alpha \ln \varphi_n(\textbf{x}_n) \Big].
\end{aligned}
\end{equation}

The remaining term, $\frac{\langle \Psi |H| \partial_\alpha \Psi \rangle}{\langle \Psi | \Psi \rangle}$, may be split up into a sum of the terms in the Hamiltonian. The corresponding expressions may be evaluated analogous to the expressions for the energy, but appended by a factor of $\partial_\alpha \ln \varphi_n$ to account for the derivative in the ket. Explicitly, the kinetic energy term may be expressed as
\begin{equation}\label{eq:KE_derivative}
\begin{aligned}
    &\frac{1}{\langle \Psi | \Psi \rangle} \Big\langle \Psi\Big| \frac{1}{2m} \int dx \frac{d\hat{\psi}^\dag(x)}{dx} \frac{d \hat{\psi} (x)}{dx} \Big| \partial_\alpha \Psi \Big\rangle \qquad \qquad \qquad \qquad \\
        &\quad = \frac{1}{\langle \Psi | \Psi \rangle} \Bigg( \Big\langle \partial_\alpha \Psi\Big| \frac{1}{2m} \int dx \frac{d\hat{\psi}^\dag(x)}{dx} \frac{d \hat{\psi} (x)}{dx} \Big| \Psi \Big\rangle \Bigg)^* \qquad \qquad \qquad \qquad\\
        &\quad = \frac{1}{\langle \Psi | \Psi \rangle} \sum_{n=0}^\infty \int d^nx \Big(\partial_\alpha \varphi^*_n(\textbf{x}_n) \cdot \tfrac{-1}{2m} \nabla^2 \varphi_n(\textbf{x}_n) \Big)^* \\
        &\quad = \mathop{\mathbb{E}}_{n\sim P_n} \mathop{\mathbb{E}}_{\textbf{x}_n \sim |\bar{\varphi}_n|^2} \Big[\big( \partial_\alpha \ln \varphi_n(\textbf{x}_n) \big)  \\
        &\qquad \qquad \qquad \quad \times\tfrac{-1}{2m} \Big(\nabla^2 \ln \varphi^*_n(\textbf{x}_n) + \big(\nabla \ln \varphi^*_n(\textbf{x}_n) \big)^2 \Big) \Big],
\end{aligned}
\end{equation}
The external and chemical potential term as
\begin{equation}\label{eq:1_body_derivative}
\begin{aligned}
    &\frac{1}{\langle \Psi | \Psi \rangle} \Big\langle \Psi \Big| \int dx \ (V(x)-\mu) \hat{\psi}^\dag(x) \hat{\psi} (x) \Big| \partial_\alpha \Psi \Big\rangle  \\
        & \qquad = \frac{1}{\langle \Psi | \Psi \rangle} \sum_{n=0}^\infty \int d^nx \ \varphi^*_n(\textbf{x}_n) \partial_\alpha \varphi_n(\textbf{x}_n) \\
        & \qquad \qquad \qquad \qquad \qquad \qquad \times \Big(\textstyle\sum_{i=1}^n (V(x_i) - \mu)\Big) \\ 
        & \qquad = \mathop{\mathbb{E}}_{n\sim P_n} \mathop{\mathbb{E}}_{\textbf{x}_n \sim |\bar{\varphi}_n|^2} \Big[\big(\partial_\alpha \ln \varphi_n (\textbf{x}_n)\big) \textstyle\sum_{i=1}^n (V(x_i)-\mu) \Big],
\end{aligned}
\end{equation}
and the two body potential as
\begin{equation}\label{eq:2_body_derivative}
\begin{aligned}
    & \frac{1}{\langle \Psi | \Psi \rangle} \Big\langle \Psi \Big| \frac{1}{2} \int dx dy \ W(x-y) \hat{\psi}^\dag(x) \hat{\psi}^\dag(y) \hat{\psi}(y) \hat{\psi}(x) \Big| \partial_\alpha \Psi \Big\rangle \\
        &\qquad = \frac{1}{\langle \Psi | \Psi \rangle} \sum_{n=0}^\infty \int d^nx \ \varphi^*_n(\textbf{x}_n) \partial_\alpha \varphi_n(\textbf{x}_n)\\
        & \qquad \qquad \qquad \qquad \qquad \qquad \times \Big(\tfrac{1}{2} \textstyle\sum_{i\neq j} W(x_i - x_j)\Big) \\ 
        &\qquad = \mathop{\mathbb{E}}_{n\sim P_n} \mathop{\mathbb{E}}_{\textbf{x}_n \sim |\bar{\varphi}_n|^2} \Big[\big(\partial_\alpha \ln \varphi_n (\textbf{x}_n)\big) \textstyle\sum_{i<j} W(x_i - x_j) \Big].
\end{aligned}
\end{equation}

Alternatively, if one desires to use the expression for kinetic energy that uses the gradient instead of the Laplacian, we may express its corresponding derivative term as 
\begin{equation}\label{eq:KE_derivative2}
\begin{aligned}
    &\frac{1}{\langle \Psi | \Psi \rangle} \Big\langle \Psi\Big| \frac{1}{2m} \int dx \frac{d\hat{\psi}^\dag(x)}{dx} \frac{d \hat{\psi} (x)}{dx} \Big| \partial_\alpha \Psi \Big\rangle \qquad \qquad \qquad \qquad \\
        &\quad = \frac{1}{\langle \Psi | \Psi \rangle} \sum_{n=0}^\infty \int d^nx \Big( \tfrac{1}{2m} \nabla \varphi_n^*(\textbf{x}_n) \cdot \nabla \partial_\alpha \varphi_n(\textbf{x}_n) \Big) \\
        &\quad = \mathop{\mathbb{E}}_{n\sim P_n} \mathop{\mathbb{E}}_{\textbf{x}_n \sim |\bar{\varphi}_n|^2} \Big[ \tfrac{1}{2m} \tfrac{\nabla \varphi_n^*(\textbf{x}_n)}{\varphi_n^*(\textbf{x}_n)} \cdot \tfrac{\nabla \partial_\alpha \varphi_n(\textbf{x}_n)}{\varphi_n(\textbf{x}_n)} \Big] \\
        &\quad = \mathop{\mathbb{E}}_{n\sim P_n} \mathop{\mathbb{E}}_{\textbf{x}_n \sim |\bar{\varphi}_n|^2} \Big[ \tfrac{1}{2m} \nabla \ln \varphi_n^*(\textbf{x}_n) \\
        &\qquad \ \  \times \Big( \big( \nabla \ln \varphi_n(\textbf{x}_n) \big) \partial_\alpha \ln \varphi_n(\textbf{x}_n) + \partial_\alpha \nabla \ln \varphi_n(\textbf{x}_n) \Big) \Big],
\end{aligned}
\end{equation}
the real part of which is 
\begin{equation}\label{eq:KE_derivative3}
\begin{aligned}
    &= \mathop{\mathbb{E}}_{n\sim P_n} \mathop{\mathbb{E}}_{\textbf{x}_n \sim |\bar{\varphi}_n|^2} \Big[ \tfrac{1}{2m} \Big( \big|\nabla \ln \varphi_n(\textbf{x}_n) \big|^2 \partial_\alpha \ln \varphi_n \\ 
    & \qquad \qquad \qquad \qquad \qquad + \tfrac{1}{2} \partial_\alpha \big|\nabla \ln \varphi_n(\textbf{x}_n) \big|^2  \Big) \Big].
\end{aligned}
\end{equation}

Lastly, for the regularized Klein-Gordon model studied in the main text, we may determine the derivative of the expectation value $\langle \lambda \int dx \hat{\psi}^\dag(x)\hat{\psi}^\dag(x) + h.c. \rangle$ by directly taking the derivative of Eq.~\eqref{eq:lam}:
\begin{equation}\label{eq:lam_derivative}
\begin{aligned}
    &\frac{\lambda}{\langle \Psi | \Psi \rangle} \sum_{n=0}^\infty \sqrt{(n+1)(n+2)} \int d^nx dx'  \\
        &\qquad \qquad \qquad \quad \times \partial_\alpha \big( \varphi_n(\textbf{x}_n) \varphi_{n+2}(\textbf{x}_n, x', x') \big)  + \text{h.c.}\\
        &\qquad = \mathop{\mathbb{E}}_{n\sim P_n} \mathop{\mathbb{E}}_{\textbf{x}_n \sim |\bar{\varphi}_n|^2} \Bigg[ 2\lambda \sqrt{(n+1)(n+2)} \\
        &\qquad \quad \times \text{Re} \Big( \frac{\int dx' \partial_\alpha \varphi_{n+2}(\textbf{x}_{n}, x', x')}{\varphi_{n}(\textbf{x}_n)} \\ 
        & \qquad \qquad \qquad + \partial_\alpha \ln\varphi_{n}(\textbf{x}_n) \frac{\int dx' \varphi_{n+2}(\textbf{x}_{n}, x', x')}{\varphi_{n}(\textbf{x}_n)} \Big) \Bigg].
\end{aligned}
\end{equation}

\subsubsection{Expressions for Energy Density}
In the section on the Lieb-Liniger model in the main text, we calculated the energy densities of our NQFS and compared them to the exact solution. Let us also show how these quantities may be evaluated.

The simplest such expression to evaluate is the particle number density, which may be written as 
\begin{equation}\label{eq:particle_number_density1}
\begin{aligned}
    &\frac{1}{\langle \Psi | \Psi \rangle} \Big\langle \Psi \Big| \hat{\psi}^\dag(x) \hat{\psi} (x) \Big| \Psi \Big\rangle \\
        &\quad = \frac{1}{\langle \Psi | \Psi \rangle} \Big\langle \Psi \Big| \int dy \delta(y-x) \hat{\psi}^\dag(y) \hat{\psi} (y) \Big| \Psi \Big\rangle. 
\end{aligned}
\end{equation}
Note that this has the form of the expectation value of an external potential $V(y) = \delta(y-x)$. Therefore, we can import Eq.~\eqref{eq:1_body} with the appropriate substitution $V(y) = \delta(x-y)$ and $\mu = 0$ to evaluate this as
\begin{equation}\label{eq:particle_number_density2}
\begin{aligned}     
    &\frac{1}{\langle \Psi | \Psi \rangle} \Big\langle \Psi \Big| \hat{\psi}^\dag(x) \hat{\psi} (x) \Big| \Psi \Big\rangle \\
        &\qquad = \frac{1}{\langle \Psi | \Psi \rangle} \sum_{n=0}^\infty \int d^nx \ |\varphi_n(\textbf{x}_n)|^2 \Big(\textstyle\sum_{i=1}^n \delta(x-x_i) \Big) \\ 
        &\qquad = \frac{1}{\langle \Psi | \Psi \rangle} \sum_{n=0}^\infty \sum_{i=1}^n \int d^{n \setminus i}x \ |\varphi_n(\textbf{x}_n|x_i=x)|^2 \\
        &\qquad =\mathop{\mathbb{E}}_{n\sim P_n}\Bigg( \sum_{i=1}^n \mathop{\mathbb{E}}_{\textbf{x}_{n\setminus i} \sim |\bar{\varphi}_{n \setminus i}|^2} \Bigg[ \frac{|\varphi_n(\textbf{x}_n|x_i=x)|^2}{\int dx_i |\varphi_n(\textbf{x}_n)|^2} \Bigg] \Bigg).
\end{aligned}
\end{equation}
Here the second expectation value is taken over the marginal distribution $|\bar{\varphi}_{n \setminus i}|^2 = \int dx_i |\bar{\varphi}_{n}|^2$, and the notation $\varphi_n(\textbf{x}_n|x_i=x)$ means that this function is evaluated at $\textbf{x}_n$ with the constraint that $x_i = x$. Samples from the marginal distribution may be obtained by sampling $\textbf{x}_n \sim |\bar{\varphi}_n|^2$ as usual, and ignoring $x_i$. In practice, we evaluate the integral in the denominator, $\int dx_i |\varphi_n(\textbf{x}_n)|^2$, with numerical integration, such as the trapezoid rule. 

Next, the kinetic energy density may be expressed as
\begin{equation}\label{eq:KE_density1}
\begin{aligned}
    &\frac{1}{\langle \Psi | \Psi \rangle} \Big\langle \Psi\Big| \frac{1}{2m} \frac{d\hat{\psi}^\dag(x)}{dx} \frac{d \hat{\psi} (x)}{dx} \Big| \Psi \Big\rangle \qquad \qquad \qquad \qquad  \\
        &\qquad \frac{1}{\langle \Psi | \Psi \rangle} \Big\langle \Psi\Big| \frac{1}{2m} \int dy \delta(y-x) \frac{d\hat{\psi}^\dag(y)}{dy} \frac{d \hat{\psi} (y)}{dy} \Big| \Psi \Big\rangle,
\end{aligned}
\end{equation}
which is again like Eq.~(\ref{eq:KE}) with an extra factor of $\delta(y-x)$. Following through with this logic, we may express this quantity as
\begin{equation}\label{eq:KE_density2}
\begin{aligned}
    &\frac{1}{\langle \Psi | \Psi \rangle} \Big\langle \Psi\Big| \frac{1}{2m} \frac{d\hat{\psi}^\dag(x)}{dx} \frac{d \hat{\psi} (x)}{dx} \Big| \Psi \Big\rangle \qquad \qquad \qquad \qquad  \\
        &\quad = \frac{1}{\langle \Psi | \Psi \rangle} \sum_{n=0}^\infty \int d^nx \ \tfrac{1}{2m} \textstyle\sum_{i=1}^n \delta(x-x_i) \big| \frac{\partial \varphi_n}{\partial x_i} (\textbf{x}_n)\big|^2 \\
        &\quad = \frac{1}{\langle \Psi | \Psi \rangle} \sum_{n=0}^\infty \sum_{i=1}^n \int d^{n\setminus i}x \  \tfrac{1}{2m} \big|\textstyle\frac{\partial \varphi_n}{\partial x_i}(\textbf{x}_n|x_i=x)\big|^2 \\
        &\quad =\mathop{\mathbb{E}}_{n\sim P_n}\Bigg( \sum_{i=1}^n \mathop{\mathbb{E}}_{\textbf{x}_{n\setminus i} \sim |\bar{\varphi}_{n \setminus i}|^2}  \Bigg[\frac{\tfrac{1}{2m} \big| \frac{\partial \varphi_n }{\partial x_i} (\textbf{x}_n|x_i=x)\big|^2}{\int dx_i |\varphi_n(\textbf{x}_n)|^2} \Bigg] \Bigg).
\end{aligned}
\end{equation}
Sampling with respect to $|\bar{\varphi}_{n \setminus i}|^2$ may be achieved as explained above, and the integral in the denominator may be calculated with numerical integration. 

Lastly, for the Lieb-Liniger model, with interaction potential $W(x-y) = 2g\delta(x-y)$, the interaction potential energy density may be expressed as
\begin{equation}\label{eq:2_body_density1}
\begin{aligned}
    & \frac{1}{\langle \Psi | \Psi \rangle} \Big\langle \Psi \Big| \frac{1}{2} \ 2g \hat{\psi}^\dag(x) \hat{\psi}^\dag(x) \hat{\psi}(x) \hat{\psi}(x) \Big| \Psi \Big\rangle = \\
    & \frac{1}{\langle \Psi | \Psi \rangle} \Big\langle \Psi \Big| \frac{1}{2} \int dy dz \ 2g \delta(y-x) \delta(z-x) \\
    & \qquad \qquad \qquad \qquad \qquad \times \hat{\psi}^\dag(y) \hat{\psi}^\dag(z) \hat{\psi}(z) \hat{\psi}(y) \Big| \Psi \Big\rangle
\end{aligned}
\end{equation}
which again has the appearance of Eq.~(\ref{eq:2_body}). Importing this equation with the substitution $W(y-z) = 2g \delta(y-x) \delta(z-x)$ and pushing through the algebra, the Lieb-Liniger interaction energy density may be expressed as 
\begin{equation}\label{eq:2_body_density2}
\begin{aligned}
    & \frac{1}{\langle \Psi | \Psi \rangle} \Big\langle \Psi \Big| g \hat{\psi}^\dag(x) \hat{\psi}^\dag(x) \hat{\psi}(x) \hat{\psi}(x) \Big| \Psi \Big\rangle \\
        &\qquad = \frac{1}{\langle \Psi | \Psi \rangle} \sum_{n=0}^\infty \int d^nx \ |\varphi_n(\textbf{x}_n)|^2 \\
        & \qquad \qquad \qquad \qquad \qquad \times \Big(\textstyle\sum_{i\neq j} g \delta(x-x_i)\delta(x-x_j) \Big) \\ 
        &\qquad = \frac{1}{\langle \Psi | \Psi \rangle} \sum_{n=0}^\infty \sum_{i\neq j} \int d^{n \setminus i,j}x \ g|\varphi_n(\textbf{x}_n|x_i,x_j = x)|^2 \\
        &\qquad =\mathop{\mathbb{E}}_{n\sim P_n}\Bigg( \sum_{i\neq j} \  \mathop{\mathbb{E}}_{\textbf{x}_{n\setminus i,j} \sim |\bar{\varphi}_{n \setminus i,j}|^2} \\
        & \qquad \qquad \qquad \qquad \qquad \qquad \Bigg[ \frac{g \cdot  |\varphi_n(\textbf{x}_n|x_i,x_j = x)|^2}{\int dx_i dx_j |\varphi_n(\textbf{x}_n)|^2} \Bigg] \Bigg).
\end{aligned}
\end{equation}
Here the second expectation value is taken over the marginal distribution $|\bar{\varphi}_{n \setminus i,j}|^2 = \int dx_i dx_j |\bar{\varphi}_{n}|^2$, and the notation $\varphi_n(\textbf{x}_n|x_i, x_j=x)$ means that this function is evaluated at $\textbf{x}_n$ with the constraints $x_i = x$ and $x_j=x$. As before, samples from this marginal distribution may be obtained by sampling $\textbf{x}_n \sim |\bar{\varphi}_n|^2$ as usual, and ignoring $x_i$ and $x_j$. And again, the integral in the denominator may be evaluated with numerical integration.

\subsubsection{Expression for One-Body Density Matrix}
In studying the Calogero-Sutherland model in the main text, we evaluated the one-body density matrix for comparison with the exact solution. On a ring of length $L$, the one-body density matrix is defined for an $n$-particle wave function as
\begin{equation}
\begin{aligned}
    g_1(x) = \frac{1}{\langle \varphi_n |\varphi_n \rangle } n \int_0^L d^{n-1}x &\varphi_n^*(x_1+x, x_2, \ldots x_n) \\
    &\qquad \ \times \varphi_n(x_1,x_2, \ldots ,x_n).
\end{aligned}
\end{equation}
This equation holds for any $x_1$, such that it may be re-expressed as~\cite{Astrakharchik_2006}
\begin{equation}
\begin{aligned}
    &\frac{1}{\langle \varphi_n |\varphi_n \rangle } \frac{n}{L} \int_0^L d^{n}x \varphi_n^*(x_1+x, x_2, \ldots x_n) \varphi_n(x_1,x_2, \ldots ,x_n) \\
    & \qquad = \mathop{\mathbb{E}}_{\textbf{x}_n \sim |\bar{\varphi}_n|^2} \Bigg[ \frac{n}{L} \frac{\varphi_n^*(x_1+x, x_2, \ldots x_n)}{\varphi_n(x_1, x_2, \ldots x_n)} \Bigg].
\end{aligned}
\end{equation}
Translating this over to the field theoretic setting, we may estimate the one-body density matrix of state $|\Psi\rangle$ by the expectation value
\begin{equation}
\begin{aligned}
    g_1(x) = \mathop{\mathbb{E}}_{n\sim P_n} \mathop{\mathbb{E}}_{\textbf{x}_n \sim |\bar{\varphi}_n|^2} \Bigg[ \frac{n}{L} \frac{\varphi_n^*(x_1+x, x_2, \ldots x_n)}{\varphi_n(x_1, x_2, \ldots x_n)} \Bigg].
\end{aligned}
\end{equation}
This is straightforwardly computed as an empirical mean over MCMC samples drawn jointly from $P_n$ and $|\bar{\varphi}_n(\textbf{x}_n)|^2$.

\section{Markov Chain Monte Carlo Sampling in Fock Space}\label{sec:MCMC}
As we showed above, a practical implementation of VMC in Fock space requires one to draw samples jointly from $P_n$ and $|\bar{\varphi}_n(\textbf{x}_n)|^2$. Here we develop an MCMC algorithm to draw these samples. This algorithm can be seen as a generalization of MCMC in the classical grand canonical ensemble~\cite{frenkel2001understanding, valleau1980primitive} to the quantum grand canonical ensemble.

We develop our algorithm from the Metropolis-Hastings algorithm, which is conventionally used to produce a chain of samples ${x_i}$ drawn from a probability distribution that it known up to its normalization. This algorithm operates by proposing a new configuration, which is then probabilistically accepted or rejected. In particular, when the current configuration of the chain is $x$, a new configuration $x'$ is proposed, and is accepted with probability
\begin{equation}
    \text{min}\left(1, \frac{\mathcal{P}(x')}{\mathcal{P}(x)} \frac{g(x|x')}{g(x'|x)} \right)
\end{equation}
where $\mathcal{P}(x)$ is the probability of configuration $x$, and $g(x'|x)$ is the probability of transitioning from $x \rightarrow x'$ under a given proposal scheme. 

We use this to generate samples from the probability distribution of a state $|\Psi\rangle$ in Fock space, i.e. $P_n$ and $|\bar{\varphi}_n(\textbf{x}_n)|^2$. First focusing on the probability $\mathcal{P}(x)$, note that for a generic $n$-particle wave function $\varphi_n(\textbf{x}_n)$, the probability of the configuration $x = \textbf{x}_n$ is 
\begin{equation}
\begin{aligned}
    &\mathcal{P}_{\text{generic}}(\textbf{x}_n) = P_n \cdot |\bar{\varphi}_n(\textbf{x}_n)|^2 d^nx \\
    & \ \ = \frac{\int{d^nx' |\varphi_n(\textbf{x}'_n)|^2}}{\sum_m \int{d^mx' |\varphi_m(\textbf{x}'_m)|^2}} \frac{|\varphi_n(\textbf{x}_n)|^2}{\int{d^nx' |\varphi_n(\textbf{x}'_n)|^2}} d^nx \\
    & \ \ = \frac{|\varphi_n(\textbf{x}_n)|^2 d^nx }{\sum_m \int{d^mx' |\varphi_m(\textbf{x}'_m)|^2}}.
\end{aligned}
\end{equation}
In this paper however, we consider bosonic states, in which a configuration $\textbf{x}_n$ is indistinguishable from all of its permutations $\tilde{\textbf{x}}_n \in \text{Sym}(\textbf{x}_n)$. The probability of the $n$-particle configuration $\textbf{x}_n$ is therefore
\begin{equation}
\begin{aligned}
    \mathcal{P}_{\text{bos}}(\textbf{x}_n) &= \sum_{\tilde{\textbf{x}}_n \in \text{Sym}(\textbf{x}_n)} \mathcal{P}_{\text{generic}}(\tilde{\textbf{x}}_n) \\
    &= \frac{n! |\varphi_n(\textbf{x}_n)|^2 d^nx }{\sum_m \int{d^mx' |\varphi_m(\textbf{x}'_m)|^2}}.
\end{aligned}
\end{equation}

Next, the transition probability $g(x'|x)$ is dictated by the proposal function, which here must propose a new configuration of particles and also allow the particle number to change. We select this function as follows. With probability $p_0$ the proposed configuration will have the same particle number, and with probability $2\cdot p_\pm$ the particle number changes, increasing or decreasing by $1$ with equal probability $p_\pm$. By normalization, $p_0 + 2p_\pm = 1 \ \Rightarrow \ p_0 = 1-2 p_\pm$, so only $p_\pm$ need be specified.

Thus, with probability $p_0$ the particle number stays the same, and we propose that the configuration change by an additive uniform random variable: $\textbf{x}_n \rightarrow \textbf{x}'_n = \textbf{x}_n + \xi_n$, where $\xi_n \sim U(-w/2, w/2)$ is a uniform random variable of width $w$. Observe that this update scheme is symmetric: $g(x|x') = g(x'|x)$, and so the acceptance probability of $\textbf{x}_n \rightarrow \textbf{x}'_n$ is
\begin{equation}
    \text{min} \left(1, \ \frac{|\varphi_n(\textbf{x}'_n)|^2}{|\varphi_n(\textbf{x}_n)|^2} \right).
\end{equation}

On the other hand, with probability $p_\pm$, the proposal function adds a new particle at position $x_{n+1}$, which is selected uniformly at random from the system $[0,L]$. Likewise, with probability $p_{-}$, the proposal removes the particle at $x_{i}$ for a uniform random $i\in \{1,...,n\}$. These proposals are not symmetric; for the addition of a particle at the random position $x_{n+1}$ (i.e. the move $\textbf{x}_n \rightarrow \textbf{x}_{n+1}$), we have $g(\textbf{x}_{n+1}|\textbf{x}_n) = p_\pm \cdot dx_{n+1}/L$, and for the removal of a random particle (i.e. the move $\textbf{x}_n \rightarrow \textbf{x}_{n-1}$), we have $g(\textbf{x}_{n-1}|\textbf{x}_n) = p_\pm \cdot 1/n$. The acceptance probability of the move $\textbf{x}_n \rightarrow \textbf{x}_{n+1}$ is then
\begin{equation}
\begin{aligned}
    &\text{min}\left(1, \ \frac{\mathcal{P}_{\text{bos}}(\textbf{x}_{n+1})}{\mathcal{P}_{\text{bos}}(\textbf{x}_{n})} \frac{g(\textbf{x}_{n}|\textbf{x}_{n+1})}{g(\textbf{x}_{n+1}|\textbf{x}_{n})} \right) = \\
    &\text{min}\left(1, \ \frac{(n+1)!}{n!} \frac{|\varphi_{n+1}(\textbf{x}_{n+1})|^2 d^{n+1}x }{|\varphi_n(\textbf{x}_{n})|^2 d^nx} \frac{p_\pm}{n+1} \frac{L}{p_\pm dx_{n+1}} \right)= \\ 
    &\text{min}\left(1, \ \frac{L |\varphi_{n+1}(\textbf{x}_{n+1})|^2}{|\varphi_n(\textbf{x}_{n})|^2}\right),
\end{aligned}
\end{equation}
and that of the move $\textbf{x}_n \rightarrow \textbf{x}_{n-1}$ is (assuming particle $n$ was removed without loss of generality)
\begin{equation}
\begin{aligned}
    &\text{min}\left(1, \ \frac{\mathcal{P}_{\text{bos}}(\textbf{x}_{n-1})}{\mathcal{P}_{\text{bos}}(\textbf{x}_{n})} \frac{g(\textbf{x}_{n}|\textbf{x}_{n-1})}{g(\textbf{x}_{n-1}|\textbf{x}_{n})} \right) = \\
    &\text{min}\left(1, \ \frac{(n-1)!}{n!} \frac{|\varphi_{n-1}(\textbf{x}_{n-1})|^2 d^{n-1}x }{|\varphi_n(\textbf{x}_{n})|^2 d^nx} \frac{p_\pm dx_n}{L} \frac{n}{p_\pm} \right)= \\ 
    &\text{min} \left(1, \ \frac{1}{L} \frac{|\varphi_{n-1}(\textbf{x}_{n-1})|^2}{|\varphi_{n}(\textbf{x}_{n})|^2} \right).
\end{aligned}
\end{equation}

Ultimately, MCMC in Fock space performs the Metropolis-Hastings algorithm with the above proposal function and acceptance ratios. This allows us to sample the particle number distribution $P_n$, and within each particle number sector sample configurations from $|\bar{\varphi}_n(\textbf{x}_n)|^2$. Therefore, the configurations of a Markov chain output by this algorithm will be jointly distributed according to $P_n$ and $|\bar{\varphi}_n (\textbf{x}_n)|^2$, and can thus be fed directly into the above equations for VMC in Fock space. 

We note that while these acceptance ratios depend on the system size $L$, this factor cancels out in our calculations. This is because the $n$-particle wave function of the NQFS defined in Eq.~\eqref{eq:NQFS_definition2} contains a multiplicative factor of $1/L^{n/2}$, which nullifies the factor of $L$ in the acceptance ratio. Hence, this factor of the system size is not detrimental to our methodology.

In our experiments, we select $p_\pm=0.25$, so that a configuration with different particle number is proposed $2\cdot p_\pm = 50\%$ of the time, thus allowing the algorithm to thoroughly sample through the entire state space (i.e. sample between different particle number sectors, and sample the $n$-particle wave function within each sector). A larger value of $p_\pm$ will force the algorithm to search more through particle number sectors, which could significantly decrease the acceptance ratio, whereas a smaller value of $p_\pm$ will limit the algorithm's ability to search through particle number sectors and may slow convergence to the optimum.

In practice, we run multiple MCMC chains in parallel on GPU, and estimate expectation values as the corresponding empirical mean over all of the samples. We estimate the uncertainty in an expectation value by binning the samples over the chains, and computing the standard deviation of the corresponding means across the chains~\cite{evertz2009computer}.

\section{Regularization}
As mentioned in the main text, we incorporate a regularization factor to assist the NQFS in optimization over particle number. This optimization can be troublesome if the magnitudes of the $n$-particle wave functions $\varphi^{\text{NQFS}}_n(\textbf{x}_n)$ vary significantly with $n$. If this magnitude increases with $n$, then the particle number distribution $P_n$ can become un-normalizable, and VMC in Fock space break down.

To keep the particle number distribution well-behaved, we multiply the $n$-particle wave function by a parameterizable regularization factor $q_n$. For this role, we select the function that behaves as a smoothed rectangular pulse and decays exponentially outside of a central window:
\begin{equation}
    q_n = \frac{1}{1+e^{-s(n-c_1)}} \frac{1}{1+e^{s(n-c_2)}}; \quad 0 \leq c_1 \leq c_2; \ \  s>0,
\end{equation}
where $(c_1+c_2)/2$ is the mean of the pulse, $c_2-c_1$ is the width of its central window, and $s$ dictates the sharpness/slope of the exponential decay away from this window. The purpose of this regularization factor is to restrict the most probable values of $n$ to the finite range $[c_1, c_2]$, thus bounding the support of the particle number distribution and rendering it normalizable. In practice, the parameters $c_1, c_2,$ and $s$ are updated via a gradient-based optimization algorithm (here, ADAM) along with the parameters of the neural networks comprising the NQFS.

In our experiments, we initialized the mean at a small value $(c_1 + c_2)/2 \sim 2$, the width at a relatively large value $c_2 - c_1 \sim 5$, and the slope parameter at a modestly small value $s \sim 1$. These choices start the optimization at a small particle number, and allow it to search through a wide range of particle number sectors during optimization.

Lastly, we would like to point out again that our framework of NQFS is generic, and that our Deep Sets construction provides one realization of it. Accordingly, there exist alternative methods for regularizing NQFS. One such approach is to directly parameterize the probability distribution of particle number. This could be achieved by using a discrete normalizing flow, which represents a probability distribution over integers (see e.g. Ref.~\cite{chen2022simulating}). Then, in the corresponding algorithm for MCMC in Fock Space, one could propose configurations of different particle numbers with an acceptance ratio governed by the probabilities of the normalizing flow. This would be an interesting avenue for a future project.

\section{Cutoff Factor}
In the main text, we studied the Lieb-Liniger model with hard walls at $x=0$ and $x=L$. These boundary conditions dictate that the $n$-particle wave function must vanish as any particle approaches the boundary. We embed these boundary conditions into our NQFS by employing a cutoff factor~\cite{sarsa2011variational, flores2008compression, sarsa2014study, laughlin2009highly} that forces the wave function to vanish at the boundaries. We multiply the $n$-particle wave function by the cutoff factor
\begin{equation}
    \left(\frac{L}{30} \right)^{-n/2} \prod_{i=1}^n \frac{x_i}{L} \Big( 1-\frac{x_i}{L} \Big),
\end{equation}
where the normalization $(\frac{L}{30})^{-n/2}$ is included for stable optimization across different particle number sectors. For simplicity we have chosen a cutoff factor that vanishes linearly as it approaches the boundary; deviations from linearity exhibited by the ground state can be learned by the remainder of the NQFS.

\section{Jastrow Factor}
Approximating the ground state of a system whose interaction potential $W(x_i-x_j)$ diverges as two particles approach each other $x_i \rightarrow x_j$ can be troublesome with VMC. To better understand this phenomenon, consider a many-body ground state $\varphi_n^0(\textbf{x}_n)$ that satisfies the eigenvalue equation
\begin{equation}
\begin{aligned}
    &H \varphi_n^0(\textbf{x}_n) = \Bigg(\frac{-1}{2m}\nabla^2 + \sum_{i} (V(x_i)-\mu) \\ 
    & \qquad \qquad \qquad \qquad \qquad \quad + \sum_{i<j} W(x_i - x_j) \Bigg)\varphi_n^0(\textbf{x}_n) \\
    &\qquad \qquad \  = E_0 \varphi_n^0(\textbf{x}_n).
\end{aligned}
\end{equation}
By definition, this is obeyed for all positions $\textbf{x}_n$. In particular, this is satisfied even as $x_i \rightarrow x_j$, which requires that the divergence of the interaction potential be cancelled by an opposing divergence in the kinetic energy. This translates to the ground state wave function having a cusp at $x_i=x_j$. If one aims to approximate the ground state with VMC, this cusp must be approximated very precisely, or else the local energy will exhibit divergences that render the computation of its expectation value inefficient or even intractable with MCMC sampling~\cite{foulkes2001quantum, kato1957eigenfunctions}. 

It is therefore instructive to directly embed this cusp into the variational ansatz, especially as the details of this cusp can be gleaned from the Hamiltonian without explicit knowledge of the exact ground state. The condition that quantifies the behavior of this cusp is known as Kato's cusp condition, which dictates that the local energy remain finite as $x_i \rightarrow x_j$~\cite{kato1957eigenfunctions, foulkes2001quantum, Pescia_2022}:
\begin{equation}
    \lim_{x_i \rightarrow x_j} E^{\text{loc}}_n (\textbf{x}_n) = \lim_{x_i \rightarrow x_j} \frac{H \varphi_n(\textbf{x}_n)}{\varphi_n(\textbf{x}_n)} < \infty.
\end{equation}
The conventional way to impose this condition into a variational ansatz is to multiply the wave function by a \emph{Jastrow factor} that exhibits the appropriate cusps at $x_i=x_j$:
\begin{equation}
    J_n(\{|x_i-x_j|\}) = \prod_{i<j} u(|x_i-x_j|),
\end{equation}
where $u(x)$ is a function chosen to satisfy Kato's cusp condition. In practice, the use of a Jastrow factor decreases the estimate of the ground state energy and its variance~\cite{foulkes2001quantum}. 

We incorporate this into our NQFS ansatz by multiplying the $n$-particle wave functions by their respective Jastrow factor $J_n(\{|x_i-x_j|\})$. Below, we illustrate the derivation of the specific Jastrow factors we used for the Lieb-Liniger model and the Calogero-Sutherland model.

\section{Lieb-Liniger Model}
The Lieb-Liniger model~\cite{Lieb_1963_1, Lieb_1963_2} describes bosons interacting with contact interaction $W(x-y) = 2g \delta(x-y)$:
\begin{equation}
\begin{aligned}
    H_{\text{LL}} = \ &\frac{1}{2m} \int dx \frac{d\hat{\psi}^\dag(x)}{dx} \frac{d \hat{\psi} (x)}{dx} - \mu \int dx \hat{\psi}^\dag(x) \hat{\psi} (x) \\
    & + g \int dx \hat{\psi}^\dag(x) \hat{\psi}^\dag(x) \hat{\psi}(x) \hat{\psi}(x).
\end{aligned}
\end{equation}
This model is often characterized by a dimensionless interaction strength $\gamma = 2mg/\varrho$, where $\varrho = n/L$ is the particle number density.

\emph{Ground State.}---
The Lieb-Liniger model is integrable and thus provides a useful numerical benchmark. In the main text, we considered a system with hard walls at $x=0$ and $x=L$, in which case the ground state can be determined via the Bethe ansatz~\cite{Gaudin_1971, Batchelor_2005}. As this Hamiltonian conserves particle number, the ground state lies in a definite particle number sector, which we denote by $n_0$. The ground state energy is given by 
\begin{equation}
    E_0 = \frac{1}{2m} \sum_{i=1}^{n_0} k_i^2 - \mu n_0 
\end{equation}
where the pseudomomenta $k_i > 0$ solve the system of transcendental equations 
\begin{equation}
    k_i L = \pi a_i + \sum_{j\neq i} \left[\arctan(\tfrac{2mg}{k_i - k_j}) + \arctan(\tfrac{2mg}{k_i + k_j}) \right],
\end{equation}
where $\{a_i\}_{i=1}^n$ are integers chosen to minimize the energy ($a_i = 1 \ \forall i$ typically suffices), and $n_0$ is the non-negative integer also chosen to minimize the energy.

The situation simplifies considerably in Tonks-Girardeau limit where $\gamma \rightarrow \infty$ (i.e. $g \rightarrow \infty$), and the model may be mapped to non-interacting spinless fermions~\cite{Girardeau_1960, PhysRevLett_Tuybens}, which allows for easy evaluation of the ground state and its properties. For instance, as we are focusing on a system with hard walls at $x=0$ and $x=L$, the ground state energy can be determined by considering non-interacting fermions in an infinite potential well:
\begin{equation}
\begin{aligned}
    E_0 &= \sum_{j=1}^{n_0} \frac{\pi^2 j^2}{2m L^2} - \mu n_0 \\
    &= \frac{\pi^2 n_0 (n_0+1)(2n_0+1)}{12m L^2} - \mu n_0,
\end{aligned}
\end{equation}
where $n_0$ is the non-negative integer that minimizes this expression, and is thus the number of particles of the ground state. By analogous considerations, the particle number density of the Tonks-Girardeau ground state is
\begin{equation}
    \sum_{j=1}^{n_0} \frac{2}{L} \sin^2\left(\frac{j \pi x}{L}\right),
\end{equation}
and kinetic energy density is
\begin{equation}
    \frac{1}{2m} \sum_{j=1}^{n_0} \frac{2}{L} \left(\frac{j\pi}{L} \right)^2\cos^2\left(\frac{j \pi x}{L}\right).
\end{equation}

\emph{Jastrow Factor.}---
Let us now determine the Jastrow factor for the Lieb-Liniger model, where $W(x-y) = 2g \delta(x-y)$. Our goal here is to ensure that the local energy remains finite as $x_i \rightarrow x_j$, which we enforce by multiplying the wave function by a Jastrow factor $\prod_{i<j} u(|x_i-x_j|)$. In this limit, the possibly divergent terms in the local energy are
\begin{equation}
    \frac{\Big(\frac{-1}{2m} \nabla^2  + 2g \delta(x_i-x_j)\Big)u(|x_i - x_j|)\varphi_{n,\text{smooth}}(\textbf{x}_n)}{u(|x_i - x_j|) \varphi_{n,\text{smooth}}(\textbf{x}_n)},
\end{equation}
where $\varphi_{n,\text{smooth}}(\textbf{x}_n)$ denotes the remaining part of the wave function that is smooth at $x_i =x_j$. Transforming to coordinates $x_{ij} := x_i-x_j$ and $x_{\text{cm}} := \frac{1}{2}(x_i + x_j)$, the remaining possibly divergent terms are
\begin{equation}\label{eq:divergent}
\begin{aligned}
    \frac{\Big(\frac{-1}{m}\frac{\partial^2}{\partial x_{ij}^2}  + 2g \delta(x)\Big)u(|x_{ij}|)}{u(|x_{ij}|)}.
\end{aligned}
\end{equation}

A simple way to ensure that Eq.~\eqref{eq:divergent} remains finite as $x_{ij} \rightarrow 0$ is to force the numerator to be 0. This amounts to the condition 
\begin{equation}
    \frac{\partial^2}{\partial x^2}u(|x_{ij}|) = 2mg\delta(x_{ij}) u(|x_{ij}|),
\end{equation}
which upon integration over a small region centered around $x_{ij}=0$ yields
\begin{equation}
\begin{aligned}
    & \big[\tfrac{\partial}{\partial x}u(|x_{ij}|)\big] \Big|^{x_{ij}=0^+}_{x_{ij}=0^-} = 2mg u(0) \ \Rightarrow \ u'(0) = mg u(0).\\
\end{aligned}    
\end{equation}
While a variety of functions $u(|x_{ij}|)$ satisfy this constraint, a simple choice is a linear function: $u(|x_{ij}|) = |x_{ij}| + \frac{1}{mg}$. Dividing by $L$ to render this dimensionless, we arrive at the following Jastrow factor for the Lieb-Liniger model:
\begin{equation}
\begin{aligned}
    J^{\text{LL}}_n(\textbf{x}_n) = \prod_{i<j}^n \left(\frac{1}{L}|x_i-x_j| + \frac{1}{mgL} \right).
\end{aligned}
\end{equation}

In practice, the magnitude of the Jastrow factor can change exponentially with $n$, which, as we mentioned in the regularization section of this Supplementary Material, can signifiantly affect the distribution over particle number and inhibit smooth optimization across particle number sectors. To correct this issue, we multiply the Jastrow factor by a normalization factor $1/\sqrt{\int d^nx |J^{\text{LL}}_n(\textbf{x}_n)|^2 }$ to keep its magnitude roughly consistent across different particle number sectors. In the Tonks-Girardeau limit ($g \rightarrow \infty $), this integral may be computed with Selberg's integral formula~\cite{selberg1944berkninger}:
\begin{equation}
\begin{aligned}
    &\int d^nx \prod_{i<j}^n \left( \frac{1}{L}|x_i - x_j|\right)^2 \\
    & \qquad \qquad \qquad \qquad  = L^n \prod_{j=0}^{n-1} \frac{\big(\Gamma(1+j))^2 \cdot \Gamma(2+j)}{\Gamma(1+n+j)}.
\end{aligned}
\end{equation}

\begin{figure}[htbp]
    \includegraphics[width=0.4\textwidth]{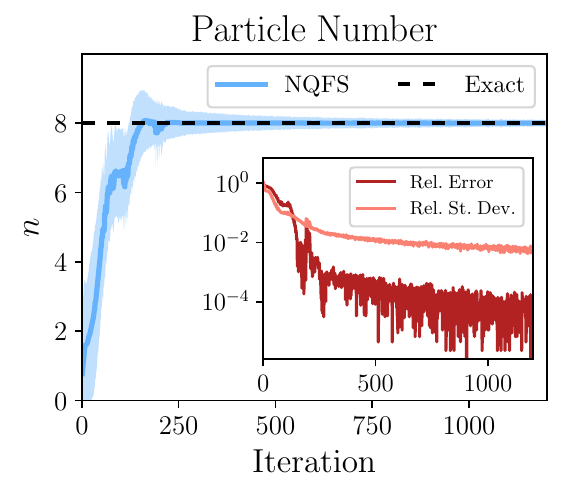}
    \caption{A plot of the particle number vs. iteration during the optimization of NQFS applied to the Lieb-Liniger model, including $\pm2$ standard deviations in light blue. The inset depicts the relative error and relative standard deviation of the particle number.} 
    \label{fig:LL_Optim_Plot}
\end{figure}

To showcase the resulting smooth optimization over particle number sectors, we show in Fig.~\ref{fig:LL_Optim_Plot} the particle number of the NQFS during optimization for the scenario studied in the main text ($L=1, \ m=1/2, \ \mu = (8.75 \pi)^2$, and $g=10^6$). This plot illustrates how the particle number converges to its exact value $n_0=8$. The relative standard deviation of the particle number decreases throughout optimization, and ultimately we find an accurate estimate $n = 8.00 \pm 0.02$.

\emph{Variance Extrapolation.}--- To further verify that NQFS accurately converges to the ground state, one can perform a technique known as variance extrapolation. Variance extrapolation is predicated on the fact that the error in the energy $\Delta E = E - E_0$ converges linearly with the variance in the energy $\sigma^2_E$ as one approaches the ground state (i.e. $\Delta E \propto \sigma^2_E$ for small $\sigma_E^2$)~\cite{Sorella_2001, Kashima_2001}. One can use this property to their advantage by making a scatter plot of $E$ vs. $\sigma^2_E$ for data collected over many iterations, and using the line of best fit to extrapolate an estimate of the exact ground state energy in the limit as the variance goes to 0. An accurate linear fit also indicates that the ground state is well approximated.

\begin{figure}[htbp]
    \includegraphics[width=0.45\textwidth]{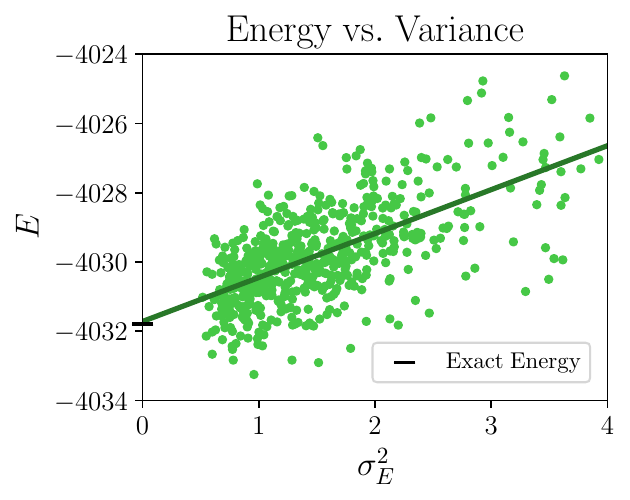}
    \caption{A scatter plot of the variational energy vs. the variance thereof, for the last 800 iterations of optimization, including the line of best fit. From this, we obtain an extrapolated energy $E=-4031.72$, which agrees exceptionally well with the exact ground state energy $E_0 = -4031.79$.}
    \label{fig:LL_VE}
\end{figure}

We illustrate variance extrapolation for the above scenario of the Lieb-Liniger model. In Fig.~\ref{fig:LL_VE}, we produce a scatter plot of $(\sigma^2_E, E)$ for data collected over many iterations of optimization, and include a linear fit to the data points with small variance $\sigma_E^2 < 4$. The results yield an extrapolated energy $E = -4031.72$, which is only $0.002\%$ greater than the exact ground state energy $E_0 = -4031.79$ and indicates that our NQFS ansatz is correctly converging to the ground state.

\section{Calogero-Sutherland Model}
The Calogero-Sutherland models describes bosons on a ring of length $L$ interacting with an inverse square sinudoidal potential $W(x-y) = g \frac{\pi^2}{L^2} \big[\sin(\frac{\pi}{L}(x-y))\big]^{-2}$~\cite{Sutherland_1971, Sutherland_1972}:
\begin{equation}
\begin{aligned}
    & H_{\text{CS}} = \ \frac{1}{2m} \int dx \frac{d\hat{\psi}^\dag(x)}{dx} \frac{d \hat{\psi} (x)}{dx} - \mu \int dx \hat{\psi}^\dag(x) \hat{\psi} (x) \\
    & \ + \frac{g \pi^2}{2 L^2} \int dx dy \hat{\psi}^\dag(x) \hat{\psi}^\dag(y) \hat{\psi}(y) \hat{\psi}(x) \big[\sin(\tfrac{\pi}{L}(x-y))\big]^{-2}
\end{aligned}
\end{equation}
Note that this interaction potential approaches an inverse square potential as $L \rightarrow \infty$.

\emph{Ground State.}---
The ground state is exactly solvable. Defining the variable $\lambda = \tfrac{1}{2}\big(1+\sqrt{1+4mg}\big)$, the ground state wave function is
\begin{equation}
    \varphi_{n_0}(\textbf{x}_{n_0}) = \prod_{i<j} \left|\sin(\frac{\pi (x_i -x_j)}{L})\right|^\lambda,
\end{equation}
and the ground state energy is
\begin{equation}\label{eq:CMS_E}
\begin{aligned}
    E_0 = \frac{\pi^2\lambda^2}{6mL^2}n_0(n_0^2-1) - \mu n_0,
\end{aligned}
\end{equation}
where $n_0$ minimizes this expression and is the number of particles of the ground state.

In the main text, we looked at the one-body density matrix 
\begin{equation}
    g_1(x) = n \int d^{n-1}x \varphi_n^*(x, x_2,\ldots , x_n) \varphi_n(0, x_2,\ldots , x_n)
\end{equation}
as a metric for the performance of NQFS. For the exact ground state, we evaluated this integral with Monte Carlo sampling.

\emph{Jastrow Factor.}---
The ground state wave function takes the form of a Jastrow factor, with $u(|x|) = \sin(\pi |x|/L)^\lambda$, which can in principle be derived from Kato's cusp condition by taking into account periodicity. Here however, we would like to construct a Jastrow factor that correctly embeds the cusp condition into the wave function, but does not exactly solve the ground state, so that the NQFS learns a nontrivial state. 

Let us construct said Jastrow factor. First, note that the interaction potential diverges as $g/(x_i-x_j)^2$ as $|x_i - x_j| \rightarrow 0$. Following logic identical to that of the Lieb-Liniger section, we find that Kato's cusp condition requires that the following quantity remains finite as $x_{ij} \rightarrow 0$:
\begin{equation}
\begin{aligned}
    \frac{\Big(\frac{-1}{m}\frac{\partial^2}{\partial x_{ij}^2}  + \frac{g}{x^2}\Big) u(|x_{ij}|) }{u(|x_{ij}|)}.
\end{aligned}
\end{equation}
We again ensure that this is finite by fixing the numerator to be $0$. Inspired by the Jastrow factor derived for the Lieb-Liniger model, we try the polynomial ansatz $u(|x_{ij}|) = |x_{ij}|^a$ for some power $a$, which satisfies the desired constraint if $a(a-1) = mg$, or equivalently $a= \frac{1}{2}(1+\sqrt{1+4mg}) = \lambda$. 

Lastly, noting also that the potential diverges as $|x_i-x_j| \rightarrow L$ because the system is on a ring, a similar analysis suggests the function $u(|x_{ij}|) = (L-|x_{ij}|)^\lambda$ for the Jastrow factor. Combining these functions into a product and dividing by $L$ to remain dimensionless, we arrive at the following Jastrow factor:
\begin{equation}
   \prod_{i<j}^n \Big( \big(\tfrac{1}{L}|x_i-x_j| \big)^\lambda \cdot \left(1-\tfrac{1}{L}|x_i-x_j| \right)^\lambda \Big).
\end{equation}

While this Jastrow factor is sufficient for the Calogero-Sutherland model, it is actually quite similar to the exact ground state, leaving little for the NQFS to learn. For this reason, we select a different Jastrow factor that is significantly dissimilar from the exact solution. To select this factor, note that the above equations really only demand that $u(|x_{ij}|)$ behave as $|x_{ij}|^\lambda$ as $x_{ij} \rightarrow 0$ (and analogously as $|x_{ij}|\rightarrow L$), such that we can replace $u(|x|)$ with any function that exhibits this asymptotic behavior. We choose a tanh() function: $u(|x_{ij}|) = \tanh(\kappa |x_{ij}|/L)^\lambda$, for a constant $\kappa$. The resulting Jastrow factor behaves like a smoothed rectangular pulse that approaches 0 as $x_{ij}\rightarrow 0$ or $x_{ij}\rightarrow L$:
\begin{equation}
\begin{aligned}
   &J^{\text{CS}}_n(\textbf{x}_n) =  \prod_{i<j}^n \tanh(\tfrac{\kappa |x_i - x_j|}{L})^\lambda \cdot \left(1 - \tanh(\tfrac{\kappa |x_i - x_j|}{L}) \right)^\lambda.
\end{aligned}
\end{equation}
This function also has the upshot that its magnitude is typically close to $1$ for large $\kappa$, such that it has a lesser impact on the magnitude of the $n$-particle wave functions and including its normalization is not crucial for $n \lesssim O(10)$. In practice, we set $\kappa = 12$.

\emph{Non-exactly Solvable Model.}---
To further emphasize the novelty and flexibility of NQFS, we study the Calogero-Sutherland model in an external potential, where the exact solution is not known. As above, this model is challenging to study with prior methods like cMPS due to its long-range interactions. We impose an external potential $V(x) = \mu\cos(4\pi x/L)$, and as in the main text, we select parameters $m = 1/2$, $g=5$, $L=5$, $\mu = 3 \cdot 7^2 \cdot \frac{\pi^2\lambda^2}{6mL^2}$. This external potential is periodic, with two minima and two maxima in the range $x \in [0,L]$, and thus we expect particles to bunch together in the minima while avoiding the maxima.

\begin{figure}[htbp]
    \includegraphics[width=0.49\textwidth]{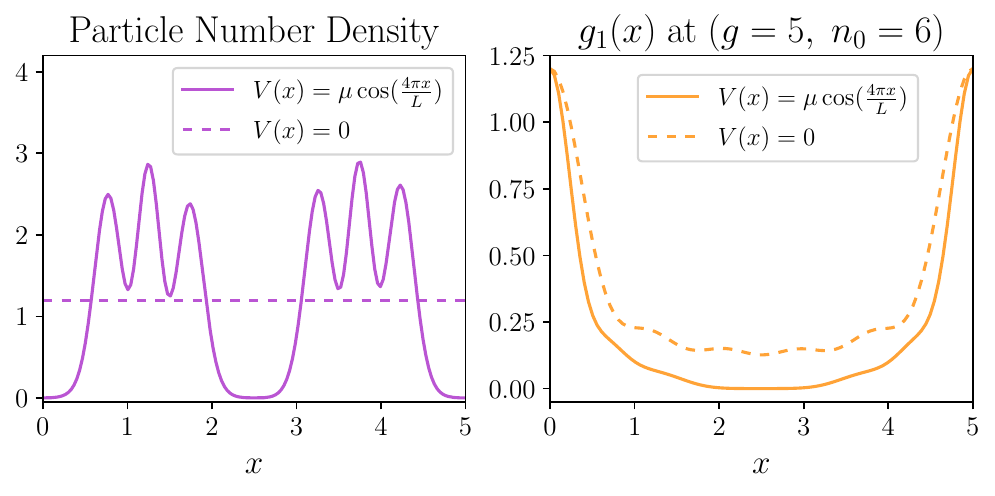}
    \caption{\textbf{Left}: A plot of the number density in the presence and absence of an external potential $V(x) = \mu \cos(4\pi x/L)$. Observe how the particles bunch together in the minima of the potential. \textbf{Right}: The one-body density matrix $g_1(x)$ in the presence and absence of an external potential.}
    \label{fig:CS_WithPotential}
\end{figure}

Upon optimizing a NQFS on this model, we find that the energy smoothly converges to $E =-563.95 \pm 0.02$, and the number of particles to $n=6.00 \pm 0.04$. In Fig.~\ref{fig:CS_WithPotential}, we illustrate the particle number density in the presence and absence of the external potential, from which we clearly see that the six particles bunch together into two clusters of three centered around the minima of $V(x)$. We also showcase the one-body density matrix $g_1(x) = n \int d^{n-1}x \varphi_n^*(x, x_2,\ldots , x_n) \varphi_n(0, x_2, \ldots , x_n)$. As the external potential separates the particles into clusters, we see that this also induces a stronger decay of correlations in the one-body density matrix. This example illustrates that NQFS produce reliable results on field theories that cannot be solved exactly.

\section{Regularized Klein-Gordon Model}
We also considered the quadratic Hamiltonian~\cite{Karanikolaou_2021, Stojevic_2015}
\begin{equation}
\begin{aligned}
    H_{\text{Quad}} = \ &\int dx \frac{d\hat{\psi}^\dag(x)}{dx} \frac{d \hat{\psi} (x)}{dx} + v \int dx \hat{\psi}^\dag(x) \hat{\psi} (x) \\
    & + \lambda \int dx \big(\hat{\psi}^\dag(x) \hat{\psi}^\dag(x) + \hat{\psi}(x) \hat{\psi}(x) \big),
\end{aligned}
\end{equation}
As we noted in the main text, this is equivalent to a regularized Klein-Gordon Hamiltonian, which can be seen by defining a non-relativistic momentum cutoff $\Lambda$ and canonically conjugate fields $\hat{\phi}(x) = \frac{1}{\sqrt{2\Lambda}}\big(\hat{\psi}(x) + \hat{\psi}^\dag(x) \big)$ and $\hat{\pi}(x) = -i\sqrt{\frac{\Lambda}{2}}\big(\hat{\psi}(x) - \hat{\psi}^\dag(x)\big)$:
\begin{equation}
\begin{aligned}
    \frac{1}{2}\int dx \ &|\hat{\pi}(x)|^2 + |\nabla \hat{\phi}(x) |^2 + m^2|\hat{\phi}(x)|^2 + \frac{1}{\Lambda^2}|\nabla \hat{\pi}(x)|^2,
\end{aligned}
\end{equation}
where $v=\frac{1}{2}(m^2 + \Lambda^2)$ and $\lambda = \frac{1}{4}(m^2 - \Lambda^2)$, and the term $\frac{1}{\Lambda^2}(\nabla \hat{\pi}(x))^2$ acts as a counterterm that regularizes the Hamiltonian.

\emph{Ground State.}---
The ground state of $H_{\text{Quad}}$ can be determined by moving to momentum space and performing a Bogoliubov transformation~\cite{Karanikolaou_2021}. In momentum space, $H_{\text{Quad}}$ becomes
\begin{equation}
\begin{aligned}
    \int \frac{dp}{2\pi} \  (p^2+v) \hat{\psi}_p^\dag \hat{\psi}_p + \lambda (\hat{\psi}_p^\dag \hat{\psi}_{-p}^\dag + \hat{\psi}_p \hat{\psi}_{-p}),
\end{aligned}
\end{equation}
where $\hat{\psi}_p = \int dx \ e^{-ipx} \hat{\psi}(x)$ is the Fourier transform of $\hat{\psi}(x)$ and obeys the commutation relation $[\psi_p, \psi_{p'}^\dag] = 2\pi \delta(p-p')$. We note that on a finite sized system of length $L$, momenta are restricted to be $p=\frac{2\pi j}{L}$ for integer $j \in \mathbb{Z}$, such that the integral is really the sum $\int \frac{dp}{2\pi} = \frac{1}{L}\sum_p$, and the delta function is $\delta(p-p') = \frac{L}{2\pi} \delta_{j,j'}$. 

The Hamiltonian in momentum space may be diagonalized by Bogoliubov transformation:
\begin{equation}
\begin{aligned}
    &\psi_p = u_p \hat{b}_p + v_p \hat{b}^\dag_{-p}, \quad |u_p|^2 - |v_p|^2 = 1,
\end{aligned}
\end{equation}
where $\hat{b}_p^\dag$, $\hat{b}_p$ are the Bogoliubov creation and annihilation operators, and $u_p$ and $v_p$ are the scalar coefficients of this decomposition. Inserting this expression and restricting to real $u_p$, $v_p$, we find that the Hamiltonian becomes diagonal if 
\begin{equation}
\begin{aligned}
     u_p v_p = \frac{-\lambda}{p^2+v} (u_p^2+v_p^2),
\end{aligned}
\end{equation}
which is solved by
\begin{equation}
\begin{aligned}
    & u_p = \sqrt{\frac{p^2+v}{2\sqrt{(p^2+v)^2-4\lambda^2}} + \frac{1}{2} } \\
    & v_p = -\sqrt{\frac{p^2+v}{2\sqrt{(p^2+v)^2-4\lambda^2}} - \frac{1}{2} }.
\end{aligned}
\end{equation}
This allows one to re-express the Hamiltonian as
\begin{equation}
\begin{aligned}
    \int \frac{dp}{2\pi}\ &\Bigg( \Big[\sqrt{(p^2+v)^2 - 4\lambda^2}\Big] \hat{b}_p^\dag \hat{b}_p \\  
    & \ \quad + \frac{2\pi \delta(0)}{2}\Big[\sqrt{(p^2+v)^2 - 4\lambda^2} - (p^2+v)\Big] \Bigg).
\end{aligned}
\end{equation}

The ground state $|\Psi_0 \rangle$ is the Bogoliubov vacuum, which obeys $b_p |\Psi_0 \rangle  = 0 \ $ for all $p$. Noting that $\delta(0)=L/(2\pi)$, we see that the ground state energy density is
\begin{equation}
    \epsilon_0 =  \frac{1}{2} \int \frac{dp}{2\pi}\Big[\sqrt{(p^2+v)^2 - 4\lambda^2} - (p^2+v)\Big]. 
\end{equation}
(Again, we emphasize that $\int \frac{dp}{2\pi} = \frac{1}{L} \sum_p$ on a finite system.)

We can also find an explicit expression for the ground state of a Bogoliubov transformation~\cite{fernandes2015lecture, demler2018strongly}: 
\begin{equation}
    |\Psi_0\rangle = e^{\int \frac{dp}{2\pi} \ \frac{v_p}{2u_p} \hat{\psi}_p^\dag \hat{\psi}_{-p}^\dag } | \Omega \rangle.
\end{equation}
It can be verified that this state obeys $b_p |\Psi_0 \rangle  = 0 \ $ for all $p$, as desired. To see this, one can use the commutation relation $[\hat{\psi}_p, \ \hat{\psi}_{p'}^{\dag n}] = 2\pi \delta(p-p') n \psi_p^{\dag (n-1)} $, which implies that $[\hat{\psi}_p, \ e^{\int \frac{dp'}{2\pi} \frac{v_{p'}}{2u_{p'}} \hat{\psi}_{p'}^\dag \hat{\psi}_{-p'}^\dag}] =  \frac{v_p}{u_p} \hat{\psi}_{-p}^\dag e^{\int \frac{dp}{2\pi} f_p \hat{\psi}_p^\dag \hat{\psi}_{-p}^\dag}$ (carefully taking into account contributions from both $p'=p$ and $p'=-p$), from which the result follows.

Let us now calculate the ground state's probability distribution of particle number, $P_n$. For ease of calculation, we will do this on a finite system, but results for an infinite system can be extrapolated by taking $L \rightarrow \infty$. First noting that $v_p$ and $u_p$ are even in $p$, the ground state may be re-expressed as 
\begin{equation}
\begin{aligned}
    |\Psi_0\rangle &= e^{\frac{1}{L} \sum_p \ \frac{v_p}{2 u_p} \hat{\psi}_p^\dag \hat{\psi}_{-p}^\dag } | \Omega \rangle \\
    &= e^{\frac{1}{L} \big[ \frac{v_0}{2u_0} \hat{\psi}_0^\dag \hat{\psi}_{0}^\dag + \sum_{p>0} \frac{v_p}{u_p} \hat{\psi}_p^\dag \hat{\psi}_{-p}^\dag \big]} | \Omega \rangle \\
    &= e^{\frac{1}{L} \frac{v_0}{2u_0} \hat{\psi}_0^\dag \hat{\psi}_{0}^\dag} \prod_{p>0} \left[ e^{\frac{1}{L} \frac{v_p}{u_p} \hat{\psi}_p^\dag \hat{\psi}_{-p}^\dag } \right]  | \Omega \rangle \\
    &= \left[ \sum_{l_0 = 0}^\infty \tfrac{1}{l_0 !} \left(\tfrac{v_0}{2Lu_0} \hat{\psi}_0^\dag \hat{\psi}_{0}^\dag\right)^{l_0} \right] \\
    & \quad \qquad \times \prod_{p>0} \left[ \sum_{l_p = 0}^\infty \tfrac{1}{l_p !} \left(\tfrac{v_p}{Lu_p} \hat{\psi}_p^\dag \hat{\psi}_{-p}^\dag\right)^{l_p} \right] |\Omega \rangle \\
    &= \left[ \sum_{l_0 = 0}^\infty \tfrac{1}{l_0 !} \left(\tfrac{v_0}{2Lu_0} \hat{\psi}_0^\dag \hat{\psi}_{0}^\dag\right)^{l_0} \right] \\
    & \quad \qquad \times \prod_{j=1}^\infty \left[ \sum_{l_j = 0}^\infty \tfrac{1}{l_j !} \left(\tfrac{v_{p_j}}{Lu_{p_j}} \hat{\psi}_{p_j}^\dag \hat{\psi}_{-{p_j}}^\dag\right)^{l_j} \right] |\Omega \rangle.
\end{aligned}
\end{equation}
where in obtaining the last line we have used the fact that the allowed momentum are $p_j := 2\pi j/L$. The first sum corresponds to a state with $2 l_0$ particles of momentum $p_0=0$; each sum in the infinite product corresponds to states that contain $l_j$ pairs of particles of opposite momenta $\pm p_j$. 

To rewrite the ground state as a sum of $n$-particle states with proper normalization, observe that the state $(\hat{\psi}_{{p}}^\dag)^n |\Omega \rangle$ has norm 
\begin{equation}
    \langle \Omega | (\hat{\psi}_{{p}})^n (\hat{\psi}_{{p}}^\dag)^n |\Omega \rangle = L^n n!,
\end{equation}
such that a normalized state with $n_j$ particles of momentum $p_j$ may be expressed as
\begin{equation}
\begin{aligned}
    &|n_1, p_1;\ n_2, p_2;\ ... \rangle =: \big|\textstyle{\prod_j} \  n_j, p_j \big\rangle  \\
    & \qquad \qquad \qquad \qquad \qquad = \prod_j \left[\frac{1}{\sqrt{L^{n_j} n_j!}} (\hat{\psi}^\dag_{p_j})^{n_j} \right] |\Omega \rangle .
\end{aligned}
\end{equation}
Therefore, making sure to treat the $p=0$ term separately, we may write the ground state as
\begin{equation}\label{eq:GS_n_representation}
\begin{aligned}
    & |\Psi_0\rangle = \sum_{l_{0}, \ l_{1}, ... = 0}^\infty \left[ \left(\tfrac{v_0}{2u_0} \right)^{l_0}  \tfrac{\sqrt{(2l_0)!}}{l_0!} \right] \prod_{j=1} \left[ \left(\tfrac{v_{p_j}}{u_{p_j}} \right)^{l_j} \right] \\
    & \qquad \qquad \qquad \qquad \qquad \quad \times \big| 2l_0, p_0;\  \textstyle{\prod_j} \  l_j, p_j;\ l_j, -p_j \big\rangle.
\end{aligned}
\end{equation}
This expression allows for a straightforward evaluation of the norm:
\begin{equation}
\begin{aligned}
    \langle \Psi_0 | \Psi_0 \rangle &= \left[ \sum_{l_0 = 0}^\infty \left(\tfrac{v_0}{2u_0} \right)^{2l_0}  \tfrac{(2l_0)!}{(l_0!)^2} \right] \times \prod_{j=1}^\infty \left[ \sum_{l_j = 0}^\infty \left(\tfrac{v_{p_j}}{u_{p_j}} \right)^{2l_j} \right] \\
    & = \tfrac{1}{\sqrt{1 - \left(\frac{v_0}{u_0}\right)^2}} \times \prod_{j=1}^\infty \Bigg[ \tfrac{1}{1-\left(\frac{v_{p_j}}{u_{p_j}} \right)^2} \Bigg] \\
    & = u_0 \times \prod_{j=1}^\infty u_{p_j}^2
\end{aligned}
\end{equation}
where we have used $u_p^2 - v_p^2 = 1$ in obtaining the last line. 

We can determine the particle number distribution from the above expression for the norm, noting that $P_n$ is the contribution to the norm from all states with $n$ particles (up to an overall normalization):
\begin{equation}
\begin{aligned}
   P_n \propto \sum_{\sum_j l_j = n/2 } \left[ \left(\tfrac{v_0}{2u_0} \right)^{2l_0}  \tfrac{(2l_0)!}{(l_0!)^2} \right] \times \prod_{j=1}^\infty \left[ \left(\tfrac{v_{p_j}}{u_{p_j}} \right)^{2l_j} \right].
\end{aligned}
\end{equation}
where the sum is over all possible combinations of $\{l_j\}$ such that $\sum_{j=0}^\infty l_j = n/2$ (remember that $l_j$ corresponds to the number of pairs of particles).

As a sanity check of these expressions, let us compute the average number of particles of the ground state $\langle n \rangle = \langle \int \frac{dp}{2\pi} \hat{\psi}_p^\dag \hat{\psi}_p \rangle$. The standard and most straightforward way to do this is to insert the Bogoliubov transformation for $\hat{\psi}_p$ and use the fact  that the ground state is the Bogoliubov vacuum:
\begin{equation}
\begin{aligned}
    \langle n \rangle & = \frac{1}{\langle \Psi_0 | \Psi_0 \rangle } \langle \Psi_0 | \int \frac{dp}{2\pi} \hat{\psi}_p^\dag \hat{\psi}_p| \Psi_0 \rangle \\
    & =\frac{1}{\langle \Psi_0 | \Psi_0 \rangle } \langle \Psi_0 | \int \frac{dp}{2\pi} \big[ u_p^2 \hat{b}_p^\dag \hat{b}_p  + u_p v_p \hat{b}_p \hat{b}_{-p} \\
    & \qquad \qquad \qquad \qquad \qquad \ + u_p v_p \hat{b}_p^\dag \hat{b}_{-p}^\dag + v_p^2 \hat{b}_{-p} \hat{b}_{-p}^\dag \big] | \Psi_0 \rangle \\
    & =\frac{1}{\langle \Psi_0 | \Psi_0 \rangle } \int \frac{dp}{2\pi} v_p^2 \langle \Psi_0|  \hat{b}_{-p} \hat{b}_{-p}^\dag   | \Psi_0 \rangle  \\
    & =\frac{1}{\langle \Psi_0 | \Psi_0 \rangle } \int \frac{dp}{2\pi} v_p^2  2\pi \delta(p=0)  \langle \Psi_0 | \Psi_0 \rangle  \\
    &= L \int \frac{dp}{2\pi} v_p^2 = \sum_p v_p^2.
\end{aligned}
\end{equation}
Alternatively, we may use our expression Eq.~\eqref{eq:GS_n_representation} for the ground state to compute this. This expression indicates that the average number of particles in the modes $\pm p_j$ is (including the appropriate normalization)
\begin{equation}
\begin{aligned}
    &\langle n_{p_j} \rangle + \langle n_{-p_j} \rangle = \frac{\sum_{l_j = 0}^\infty 2l_j \left(\tfrac{v_{p_j}}{u_{p_j}} \right)^{2l_j}}{\left[1 - \left(\tfrac{v_{p_j}}{u_{p_j}} \right)^2 \right]^{-1}} \\
    & \qquad \qquad = \frac{2 \left(\tfrac{v_{p_j}}{u_{p_j}} \right)^2}{1 - \left(\tfrac{v_{p_j}}{u_{p_j}} \right)^2} = 2 v_{p_j}^2
\end{aligned}
\end{equation}
for $p_j > 0$, and 
\begin{equation}
    \langle n_{0} \rangle = \frac{\sum_{l_0 = 0}^\infty 2l_0 \left(\tfrac{v_0}{2u_0} \right)^{2l_0}  \tfrac{(2l_0)!}{(l_0!)^2} }{\left[1 - \left(\tfrac{v_{0}}{u_{0}} \right)^2\right]^{-1/2}} = \frac{ \left(\tfrac{v_{0}}{u_{0}} \right)^2}{1 - \left(\tfrac{v_{0}}{u_{0}} \right)^2} = v_{0}^2
\end{equation}
for $p=0$. Taking into account all momentum modes, we thus find a mean number of particles
\begin{equation}
    \langle n \rangle  = \langle n_0 \rangle + \sum_{p>0} \big( \langle n_{p} \rangle  + \langle n_{-p} \rangle \big) = 
    v_0^2 + \sum_{p>0} 2 v_p^2 = \sum_p v_p^2,
\end{equation}
which is indeed consistent with the prior approach.

\emph{Cusp Condition.}---
We mentioned in the main text that this model exhibits a peculiar cusp condition, which we do not explicitly embed into the NQFS ansatz and as a result can hinder the performance of NQFS on this model. This cusp condition arises from the local energy of the term $\int dx' \hat{\psi}^\dag (x') \hat{\psi}^\dag (x') + h.c. $, which evaluates to
\begin{equation}
\begin{aligned}
    &\frac{\langle \textbf{x}_n | \int dx' \hat{\psi}^\dag (x') \hat{\psi}^\dag (x') | \Psi \rangle }{\langle \textbf{x}_n | \varphi_n \rangle }  + h.c. = \\ 
    & \qquad \frac{\sqrt{n(n-1)}}{\varphi_n(\textbf{x}_{n})}  \frac{1}{n!}\sum_{\tilde{\textbf{x}}_n \in \text{Sym}(\textbf{x}_n)}  \varphi_{n-2}(\textbf{x}'_{n-2}) \delta(x'_{n-1} - x'_n) \\
    & \qquad + h.c.
\end{aligned}
\end{equation}
This expression diverges at $x_i = x_j$ due to the presence of the delta function. 

As per Kato's cusp condition, we want the local energy to remain finite, such that its expectation value can be accurately estimated with MCMC sampling. However, it is nontrivial to enforce the above local energy to be finite, as it depends on both the $n$ and $(n-2)$-particle wave functions, and cannot be corrected with a simple Jastrow factor. 

Speculating for now, one approach could be to cancel the divergence at $x_i=x_j$ with an equivalent divergence in $\varphi_n(\textbf{x}_n)$ in the denominator. It may also be easier to analyze and fix this cusp condition by working in momentum space rather than position space. Resolving the conundrum of this cusp condition, and other cusp conditions imposed by QFT Hamiltonians, would make for an interesting study in the future.
$ $\\

\section{Implementation Details}
In the applications of the main text, we modelled the neural networks comprising the Deep Sets of the NQFS as feed-forward neural networks of width $100$ and depth $3$. This implementation was done in Pytorch~\cite{paszke2019pytorch}. MCMC sampling was performed in parallel across 100 chains, each of length 800-1200, with probability of changing particle number $p_\pm = 0.25$. ADAM optimization~\cite{kingma2014adam} was performed with the standard parameter values $\beta_1 = 0.9$ and $\beta_2 = 0.999$ We used learning rate $\eta \in [10^{-4}, 10^{-3}]$ for the neural networks, and $\eta_q \in [10^{-2}, 10^{-1}] $ for the parameters of the regularization factor $q_n$. Each of the optimizations carried out in the applications section of the main text took $\sim 3$ hours on a NVIDIA V100 GPU.

\end{document}